\documentclass[aps,prd,showpacs,floatfix,twocolumn,nofootinbib,superscriptaddress]{revtex4-1} %
\usepackage{dcolumn}% Align table columns on decimal point
\usepackage{amsmath,amsfonts,amssymb,mathrsfs,times,bm}
\usepackage{extarrows}
\usepackage{graphicx}
\usepackage{float}
\usepackage{booktabs}
\usepackage{graphicx,epstopdf}% Include figure files
\usepackage{dcolumn}% Align table columns on decimal point
\usepackage{exscale}
\usepackage{relsize}
\usepackage{latexsym}
\usepackage{longtable}
\usepackage[colorlinks=true,breaklinks=true,linkcolor=blue,citecolor=blue,urlcolor=blue]{hyperref}

\newcommand{\nn}{\nonumber}

\newcommand{\orcid}[1]{\href{https://orcid.org/#1}{\includegraphics[scale=0.035]{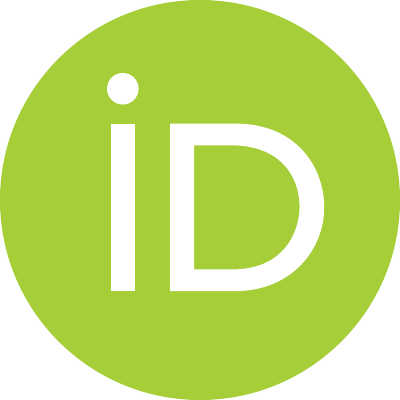}}}

\begin{document}

\title{Gravitational lensing of massive particles by a black-bounce-Schwarzschild black hole}

\author{Guansheng He\hspace*{0.6pt}\orcid{0000-0002-6145-0449}\hspace*{0.8pt}}
\email{hgs@usc.edu.cn}
\affiliation{School of Mathematics and Physics, University of South China, Hengyang 421001, China}
\affiliation{Purple Mountain Observatory, Chinese Academy of Sciences, Nanjing 210023, China}
\affiliation{Hunan Key Laboratory of Mathematical Modeling and Scientific Computing, Hengyang 421001, China}
\author{Yi Xie\hspace*{0.6pt}\orcid{0000-0003-3413-7032}\hspace*{0.8pt}}
\email{yixie@pmo.ac.cn}
\affiliation{Purple Mountain Observatory, Chinese Academy of Sciences, Nanjing 210023, China}
\author{Chunhua Jiang}
\affiliation{School of Mathematics and Physics, University of South China, Hengyang 421001, China}
\affiliation{Hunan Key Laboratory of Mathematical Modeling and Scientific Computing, Hengyang 421001, China}
\author{Wenbin Lin\hspace*{0.6pt}\orcid{0000-0002-4282-066X}\hspace*{0.8pt}}
\email{Corresponding author. lwb@usc.edu.cn}
\affiliation{School of Mathematics and Physics, University of South China, Hengyang 421001, China}
\affiliation{Hunan Key Laboratory of Mathematical Modeling and Scientific Computing, Hengyang 421001, China}
\affiliation{School of Physical Science and Technology, Southwest Jiaotong University, Chengdu 610031, China}

\date{\today}

\begin{abstract}
We investigate in detail the weak-field gravitational lensing of a relativistic neutral massive particle induced by a regular black-bounce-Schwarzschild black hole proposed by Simpson and Visser. Starting with the calculation of the gravitational deflection of the massive particle up to the third post-Minkowskian order, the Virbhadra-Ellis lens equation is solved perturbatively beyond the weak-deflection limit to achieve the expressions for the lensing observables of the primary and secondary images of a point-like particle source. The main observables contain not only the positions, the flux magnifications, and the gravitational time delays of the individual images, but also the positional relations, the magnification relations (including the total magnification), the magnification centroid, and the differential time delay. We then discuss the velocity-induced effects originated from the deviation of the particle's initial velocity from the speed of light on the black-bounce-Schwarzschild lensing observables of the images of a point-like light source, and the effects induced by the bounce parameter of the spacetime on the measurable image properties of Schwarzschild lensing of the massive particle. As an application of the results, we model the supermassive black hole in the Galactic Center (i.e., Sgr A$^{\ast}$) as the lens, and focus on evaluating the possibilities to detect the new velocity-induced and bounce-induced effects on the practical lensing observables and analyzing the dependence of these effects on the parameters.

\end{abstract}

%\pacs{04.20.-q, 95.30.Sf, 98.62.Sb}

\maketitle

\section{Introduction}
Since the first observation of the gravitational deflection of light by the Sun in 1919~\cite{DED1920}, gravitational lensing has developed into one of the most powerful tools in modern astrophysics. The substantial and extensive impact of gravitational lensing can be revealed distinctly by its fruitful applications, which include the determination of the Hubble constant~\cite{Refsdal1964}, the weak- or strong-field tests of Einstein's general relativity (GR)~\cite{Schiff1960,We1972,Reyes2010,Collett2018,CLCS2020,MWS2020,LLQZ2022} and alternative theories of gravity~\cite{KP2005,KP2006a,KP2006b,ZLBD2007,NKYS2013,Liu2016,LWLM2017,MWS2020b}, the probe of the distribution of dark and visible matter~\cite{BN1992,Wambsganss1998,JS2019}, the restriction on the mass of a neutrino~\cite{LM2019,MLMHBN2019} or a graviton~\cite{CJMM2004,FLBPZ2017,CL2021}, and so on.
Hence, it has been the subject of a large number of studies (see, for instance,~\cite{WCW1979,VE2000,BCIS2001,ERT2002,Bozza2002,Sereno2004,WS2004,CJ2009,WLFY2012,TKNA2014,
ZX2016,TG2017,LX2019E,JBO2019,KIG2020a,IKG2020b,ZX2020a,Frost2023}), with a great encouragement from the discovery of the first doubly imaged quasar~\cite{WCW1979}. Since only messengers of the electromagnetic force can be used to provide information about our physical Universe in traditional multiwavelength astronomy, most of the previous works focused on gravitational lens effects of light in a variety of spacetimes within the framework of GR or alternative theories of gravity. For example, Virbhadra and Ellis~\cite{VE2000} investigated the strong-deflection gravitational lensing of light in the Schwarzschild black hole spacetime, and proposed a popular lens equation which was adopted to study the properties of the primary, secondary, and relativistic images. Based on the strong field limit method developed by Bozza~\cite{BCIS2001,Bozza2002}, the effect of gravitational lensing caused by a deformed Ho\v{r}ava-Lifshitz black hole was also considered~\cite{CJ2009}.

Predicted by GR and verified directly by the recent observations of the Event Horizon Telescope collaboration~\cite{EHT2019a,EHT2019b} and the LIGO-Virgo collaboration~\cite{Abbott2016a,Abbott2016b}, a black hole is a solution of the equations of gravity, and serves as one of the most important gravitational sources in our Universe, given that it is widely regarded as a natural and ideal laboratory for probing fundamental physics and gravity theories. Compared with their non-regular counterparts, regular black holes are intuitively attractive to the relativity community due to the nonsingular nature.
They were explored historically from the phenomenological properties and other aspects (see~\cite{ES2011,ES2013,GN2016,FZ2017,ATSSA2017,Liang2017,ZX2017,JOSVG2018,ZJ2018,LMWX2019,Ovgun2019,Bargue2020,GD2020,GD2021,RS2023,DJCC2023}, and references therein) and many theoretical proposals with respect to regular black holes were thus performed (see, e.g.,~\cite{Ba1968,RB1983,Borde1997,Hayward2006,BM2013,FW2016,HY2019}). Recently, Simpson and Visser~\cite{SV2019} constructed a novel regular black hole spacetime, the so-called black-bounce-Schwarzschild spacetime. This geometry is especially interesting, since it (i) stands for minimal violence to the ordinary Schwarzschild spacetime in some sense, and (ii) describes a Schwarzschild black hole, a regular black hole with a one-way spacelike throat, a one-way wormhole with a null throat, or a Morris-Thorne traversable wormhole, by adjusting the bounce parameter. In recent years, the weak- and strong-field gravitational lens effects of light signals in the black-bounce-Schwarzschild spacetime or in the generalized versions~\cite{SMV2019,LSV2020,FLMSV2021,LRSSV2021,MFL2021,SPPS2021,JCCH2021,YH2021} of this geometry have been of growing interest~\cite{NPPS2020,Ovgun2020,Tsuka2021a,Tsuka2021b,CX2021,IKG2021,GB2022,ZX2022,Tsuka2022,GLM2022,CGB2023,Vagnozzi2023,GL2023,Shaikh2023,JAPO2023,SAAK2023}. For instance, Nascimento \emph{et al.}~\cite{NPPS2020} discussed firstly the strong-field gravitational lensing of light in the Simpson-Visser black-bounce spacetime. Zhang and Xie~\cite{ZX2022} studied in detail the null gravitational lens effects in the black-bounce-Reissner-Nordstr\"{o}m spacetime in the weak- and strong-field limits, respectively, and estimated the detectability of the lensing observables by modeling two typical astrophysical black holes as the lens respectively.

As is known, there are three fundamental forces of nature (i.e., the gravitational, the weak and the strong forces), besides the electromagnetic force. The non-photonic messengers of those three forces, such as neutrinos~\cite{BV1992} and cosmic-ray particles~\cite{Ande1932,Rossi1941}, can also provide valuable and specific information about our physical Universe individually or collectively~\cite{MFHM2019,Huerta2019}. However, to our knowledge, there have been few works devoted to the consideration of the propagation of timelike particles in the mentioned black-bounce-Schwarzschild spacetime or in its charged version (see, e.g.,~\cite{ZX2020b,ZX2022b,VRSA2023,MBRAS2024}), and the weak- or strong-field gravitational lensing of massive particles in either of those spacetimes has not been discussed in the literatures so far. Actually, with the coming of the era of multi-messenger astronomy, it is worth performing a full theoretical analysis of the gravitational lens effect of a massive particle induced by a black-bounce-Schwarzschild black hole, for which four reasons are responsible. Firstly, the decrease of the initial velocity of a test particle at infinity results in the increase of its total bending angle for a given gravitational lens~\cite{Silverman1980,AR2002,AR2003,WS2004}, which may make the lensing effect of a massive particle caused by the gravitational system more noticeable than the optical counterpart under the same conditions. This feature of gravitational lensing of timelike particles is of great significance, since it (i) brings to a larger opportunity for observing gravitational lensing events, and (ii) makes the consideration of the first-, second-, and even higher-order contributions to the lensing observables of the images non-trivial. It thus serves as one of the main motivations for studying the gravitational lensing of massive particles in various spacetimes (including the Simpson-Visser spacetime)~\cite{AP2004,BSN2007,PNH2014,Tsupko2014,HL2016,LYJ2016,HL2017b,BT2017,CG2018,J2018,CGV2019,CGJ2019,JBGA2019,LZLH2019,PJ2019,JL2019,LJ2020b,LHZ2020,LJ2020a,LO2020,JH2021,LLJ2021,HL2022,
HSC2023,HC2023,HCL2024}. Secondly, we know the deviation of the initial velocity of a timelike particle from the speed of light leads to the deviation of a timelike geodesic from a lightlike geodesic and affects the observable lensing properties subsequently. This deviation leads to the so-called velocity-induced effect~\cite{WS2004,HL2014,LYJ2016} on the lensing observables, which represents a crucial difference between the gravitational lensing phenomena of light and massive particles in any curved background spacetime. Given that the velocity effect on an image observable in or beyond the weak-deflection limit may be so evident that its value can be much larger than that of the corresponding null observable, more and more attentions have thus been paid to it in the last decades (see, for example,~\cite{WS2004,HL2014,LYJ2016,PJ2019,JL2019,LJ2020b,HL2022}). Especially, Wucknitz and Sperhake~\cite{WS2004} studied the mentioned velocity effect resulted from the deviation, as well as the influence of the translational motion of the lens, on the first-order gravitational deflection of light via a Lorentz boosting technique. Recently, the leading-order velocity effects on the deflection angle, the image positions, and the magnifications for both ultrarelativistic and nonrelativistic particles in Schwarzschild geometry were discussed in the weak- and strong-field limits~\cite{LYJ2016}. More recently, the weak-field gravitational lensing of a relativistic massive particle induced by a Kerr-Newman black hole was also investigated detailedly in~\cite{HL2022}, where the expressions of the velocity effects on the lensing observables of the images were obtained and the detectability of them was also evaluated. However, it should be fair to mention that further work is needed regarding the issue of velocity effects, and we can expect that it should be interesting to consider systematically the velocity effects on the lensing observables in a regular spacetime such as the black-bounce-Schwarzschild spacetime, which has not yet been reported. A third reason lies in that it is possible to provide additional information about the characteristics of the lens and the particle source or to place supplementary constraints on the spacetime parameters of the black hole, via the probe of the gravitational lens effects of timelike particles in Simpson-Visser geometry or in other regular spacetimes. It may also promote the development of mono-messenger or joint multi-messenger astronomical observations~\cite{MFHM2019,Keivani2018,IceCube2018}, considering that all of the messengers emitted by an astrophysical source may experience different geometrical and physical processes before reaching their detectors. Finally, it is widely known that rapid improvement of the high-accuracy astronomical instruments and techniques, such as the very long baseline interferometry (VLBI) techniques, has been achieved in the past decades. Current surveys and forthcoming telescope networks for multiwavelength observations aim at an astrometric precision of $1\!\sim\!10$ microarcseconds ($\mu$as) or better~\cite{Perryman2001,Prusti2016,SN2009,Reid2009,Chen2014,Malbet2012,RH2014,Malbet2014,Trippe2010,ZRMZBDX2013,Murphy2018,RD2020,Brown2021,LXLWBLYHL2022,LXBLLLH2022}. For instance, a record parallax precision of $\pm3\mu$as via the Very Long Baseline Array was reported in 2013~\cite{ZRMZBDX2013}. The Square Kilometre Array~\cite{BBGKW2015,LXLWBLYHL2022} and other next-generation radio observatories (see, e.g.,~\cite{Murphy2018,RD2020}) aim at an angular accuracy of about $1\mu$as. And the planned Nearby Earth Astrometric Telescope (NEAT) mission~\cite{Malbet2012,Malbet2014} is working towards an unprecedented space-borne astrometric accuracy of $0.05\,\mu$as. Compared with them, current astronomical detectors or instruments for observing massive particles or multi-messengers have much lower angular precisions which are at the level of one degree or better~\cite{Aab2014,Bartoli2019,Albert2020}. Exemplarily, the angular resolution of the ANTARES neutrino telescope is about $0.59^{\circ}\pm0.10^{\circ}$ for downward-going muons~\cite{Albert2020}. Moreover, the current photometric precision is at the level of about $10\,\mu$mag or better~\cite{Koch2010,BK2018,Kurtz2005}. For example, the Kepler Mission reached an extreme photometric precision of a few $\mu$mag~\cite{Koch2010,BK2018}. It ended prematurely in 2013, and was renamed as the K2 mission with a slightly lower photometric precision~\cite{VJ2014,AHILR2015,Huber2016}. Additionally, the IceCube neutrino observatory has reached a time resolution of better than 3ns~\cite{IceCube2006,AH2018}, while the VLBI techniques~\cite{Burke1969,Rogers1970,TMS1986,Hirabayashi1998,Ma1998,HKS2000,K2001} in measuring the differential time delay even have a precision of a few picoseconds (ps) currently. The proposed delay precision of the next-generation VLBI system is 4ps~\cite{Petra2009,SB2012,Niell2018}. We can expect that high-precision astronomical techniques nowadays or in near future may provide us the chance to detect the first-, second-, and even higher-order contributions to the lensing observables in the black-bounce-Schwarzschild spacetime or in other regular spacetimes, and to measure the velocity-induced effects on the properties of the lensed images.

In this work, we apply the procedure of~\cite{KP2005,HL2022,CX2021,ZX2022} to the study of the gravitational lensing of a relativistic neutral massive particle in a black-bounce-Schwarzschild black hole spacetime beyond the weak-deflection limit. We first base on the standard perturbation analysis~\cite{Murdock1999} to derive the equatorial gravitational deflection angle of the timelike particle induced by the regular black hole in the third post-Minkowskian (PM) approximation, which is employed for solving the Virbhadra-Ellis lens equation~\cite{VE2000}. Then we obtain the weak-field expressions for the lensing observables of the primary and secondary images of a point-like particle source. The observable lensing quantities mainly contain the positions, the magnifications, and the gravitational time delays of the individual images, along with the sum and difference relations for the positions or the magnifications of the images (including the total magnification), the magnification centroid, and the differential time delay between the images. The velocity-induced effects caused by the mentioned deviation on the black-bounce-Schwarzschild lensing observables of the images of a point-like light source, together with the effects resulted from the bounce parameter of the geometry on the Schwarzschild lensing properties of the images of the particle source, are discussed subsequently. Eventually, the supermassive black hole in the Galactic Center, Sgr A$^{\ast}$~\cite{EG1997,NMGPG1998,BS1999}, is assumed to be the lens as an application of our results. Under this astrophysical scenario, we estimate the possibilities to detect the velocity-induced and bounce-induced effects on the practical lensing observables, and probe the dependence of these two effects on the source position, the particle's initial velocity, and the bounce parameter.

This paper is organized as follows. In Section~\ref{sect2} we present the assumptions and notations of this work. Section~\ref{sect3} begins with a short description of the black-bounce spacetime proposed by Simpson and Visser, and is then devoted to calculating the gravitational deflection of a relativistic neutral massive particle due to a regular black-bounce-Schwarzschild black hole in the 3PM approximation. We solve the Virbhadra-Ellis lens equation, and adopt a standard perturbation method to achieve the analytical expressions of the lensing observables of the images of a point-like massive-particle source in the weak-field limit in Section~\ref{sect4}. Section~\ref{sect5} gives a discussion of the velocity-induced effects on the observable image properties in the black-bounce-Schwarzschild black hole lensing of light, and the bounce-induced effects on the Schwarzschild lensing observables of the images of the particle source. In Section~\ref{sect6}, we present an application of our formulae by assuming the Galactic Center black hole, Sgr A$^{\ast}$, to be the black-bounce-Schwarzschild lens, and concentrate on analyzing the possibilities to detect the new velocity-induced and bounce-induced effects on the practical lensing observables and probing the dependence of these effects on the parameters. Finally, a summary and a brief discussion of the results are presented in Section~\ref{sect7}. Geometrized units $(G=c=1)$ and the metric signature $(+,~-,~-,~-)$ are employed throughout this paper. Greek indices run over $0,~1,~2$, and $3$ conventionally, unless indicated otherwise.

\begin{figure}[t]
\centering
\includegraphics[width=\linewidth]{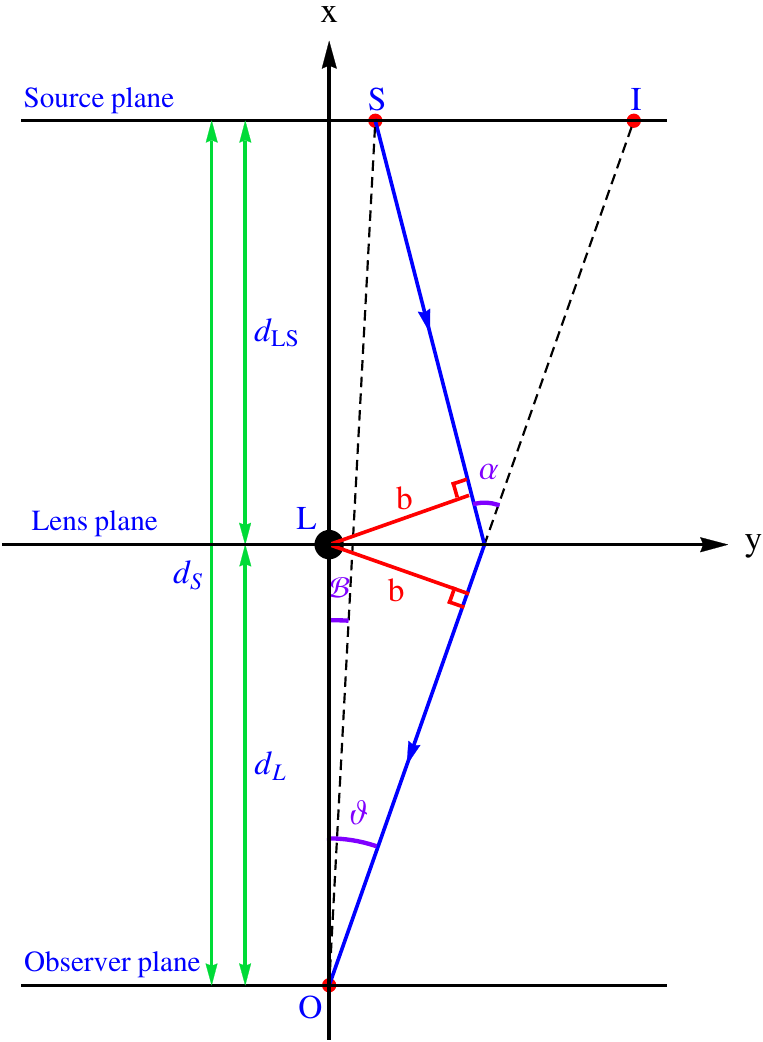}
\caption{Geometry for lensing by a black-bounce-Schwarzschild black hole. The barycenter of the black hole is located at the origin of a two-dimensional Cartesian coordinate system $(x,~y)$, the $x$ axis of which is regarded as the optic axis joining the observer and the lens. The propagating trajectory of a test particle is approximated by its two asymptotes (blue) in the thin lens approximation~\cite{Wambsganss1998}. }   \label{Figure1}
\end{figure}
\section{Basic assumptions} \label{sect2}
Let $w$ be the initial velocity of a relativistic massive (or massless) particle which is emitted by a point source, deflected by a point lens, and received by a distant observer without looping around the central body. Due to the relativistic characteristic of $w$, we may assume its rough lower limit to be about $0.05$ for the convenience of discussion and have $0.05\lesssim w\leq1$, where $w=1$ corresponds to the special case of light. Thus, according to Fermat's principle~\cite{Sch1985}, only the positive-parity primary image and negative-parity secondary image are considered in our scenario within the weak-field approximation. The lensing geometry by a black-bounce-Schwarzschild black hole is presented in Fig.~\ref{Figure1}. The observer, lens, source, and image are denoted by $O,~L,~S$, and $I$, respectively, and they are situated in the lens' equatorial plane (i.e., $x$-$y$ plane). $d_{L}$, $d_{S}$, and $d_{LS}$ are the observer-lens, observer-source, and lens-source angular diameter distances respectively. $\mathcal{B}$ and $\vartheta$ denote respectively the angular source and image positions. The gravitational deflection angle of the massive particle and the impact parameter are represented by $\alpha$ and $b~(=d_L\sin\vartheta)$, respectively. Although these angular quantities appear small in the weak-field approximation, our discussion is performed beyond the weak-deflection limit (as done in~\cite{KP2005,KP2006a}), since we aim to consider corrections to the standard weak-deflection lensing features.

For convenience, we assume the angular position of a lensed image to be always positive~\cite{KP2005}. Hence, the angular source position $\mathcal{B}$ is positive for the case that the image is on the same side of the lens (or the optic axis) as the source, and is negative when the image is on the opposite side. Moreover, our consideration of the lensing effect is assumed to perform under the thin lens approximation~\cite{Wambsganss1998,WS2004} which guarantees that all the deflection action takes place in a cosmologically small region around the lens.

\section{Weak-field gravitational deflection of massive particles} \label{sect3}
We consider the equatorial gravitational deflection of a relativistic massive particle caused by a black-bounce-Schwarzschild black hole in the weak-field limit.

\subsection{The black-bounce-Schwarzschild spacetime}
The line element for the geometry of a static spherically symmetric regular black-bounce-Schwarzschild black hole with a one-way spacelike throat takes the form in standard coordinates $(t,~r,~\theta,~\varphi)$~\cite{SV2019,SMV2019,LRSSV2021}
\begin{eqnarray}
&&\nn ds^2=\left(1-\frac{2M}{\sqrt{r^2+\eta^2}}\right)dt^2-\frac{1}{1-\frac{2M}{\sqrt{r^2+\eta^2}}}dr^2 \\
&&\hspace*{30pt} -\left(r^2+\eta^2\right)\left(d\theta^2+\sin^2\theta d\varphi^2\right)~,  \label{Metric}
\end{eqnarray}
which describes a bounce into a future universe. Here, $M$ denotes the rest mass of the black hole, and $\eta~(0<\eta<2M)$ is a bounce parameter, which takes the dimension of length and is responsible for the regularisation
of the spacetime metric. Actually, Eq.~\eqref{Metric} can also represent the geometry of a different celestial body via adjusting the non-negative parameter $\eta$.
It describes (i) a one-way wormhole geometry with a null throat when $\eta=2M$, (ii) a traversable wormhole geometry in the Morris-Thorne sense if $\eta>2M$,
(iii) the Schwarzschild spacetime when $\eta=0$, and (iv) the Ellis-Bronnikov wormhole spacetime~\cite{Ellis1973,Bronnikov1973} if $M=0$ and $\eta\neq0$. In this paper we focus on the lensing effect in the black-bounce-Schwarzschild black hole spacetime which is everywhere regular.

\subsection{Equations of motion}
The geodesic equation of a test body is the Euler-Lagrangian equation with the Lagrangian $\mathcal{L}=\frac{1}{2}g_{\mu\nu}\dot{x}^\mu\dot{x}^\nu$ for a given gravitational system~\cite{St1984,WS2004}, where the dot denotes differentiation with respect to proper time. For a massive particle propagating in the equatorial plane ($\theta=\pi/2$) of the regular black hole, we have
\begin{eqnarray}
2\mathcal{L}=\!\left(\!1\!-\!\frac{2M}{\sqrt{r^2\!+\!\eta^2}}\!\right)\!\dot{t}^{\hspace*{0.5pt}2}
\!-\!\frac{1}{1\!-\!\frac{2M}{\sqrt{r^2+\eta^2}}}\,\dot{r}^2\!-\!\left(r^2\!+\!\eta^2\right)\!\dot{\varphi}^2~.~~~~~\label{L1} \\ \nn
\end{eqnarray}
The conserved orbital energy and angular momentum per unit mass are thus given respectively by~\cite{AR2002,AR2003}:
\begin{eqnarray}
&& \bar{E}=\left(1-\frac{2M}{\sqrt{r^2+\eta^2}}\right)\dot{t}=\frac{1}{\sqrt{1-w^2}}~, \label{C1}  \\
&& \bar{L}=\left(r^2+\eta^2\right)\dot{\varphi}=\frac{w\,b}{\sqrt{1-w^2}}~,~~~~\label{C2}  \\ \nn
\end{eqnarray}
where the impact parameter $b$ is defined via the relation $\bar{L}/\bar{E}\equiv w\,b$~\cite{Tsupko2014,LYJ2016,CG2018,JBGA2019,2023IR}. It follows from Eqs.~\eqref{C1} - \eqref{C2} and the consideration of the timelike orbit $2\mathcal{L}=1$ that
\begin{equation}
\dot{r}^2=\bar{E}^2-\left(1-\frac{2M}{\sqrt{r^2+\eta^2}}\right)\left(1+\frac{\bar{L}^2}{r^2+\eta^2}\right)~,  \label{dot-r}
\end{equation}
in agreement with Eq.\,(9) of~\cite{ZTX2020}.

\subsection{Weak-field gravitational deflection of a massive particle}
We now compute the equatorial gravitational deflection angle of a massive particle in the black-bounce-Schwarzschild black hole spacetime up to the 3PM order,
via a classical approach~\cite{KP2005,We1972,HL2022}.

In the weak-field and small-angle approximation ($M/b\ll1$), we expect that the deflection angle has the power-series expansion within the PM approximation scheme
\begin{equation}
\alpha=\sum\limits_{i=1}^3 C_i\left(\frac{M}{b}\right)^i+\mathcal{O}(M^4)~, \label{alpha-general}
\end{equation}
with $C_i$ being the unknown functions of $w,~M$, and $\eta$. In order to derive its explicit expression, on the one hand we have to know the weak-field relation between $b$ and $r_0$, where $r_0$ is the distance of closest approach to the central body and satisfies the condition $M/r_0\ll1$ to guarantee a weak field. The fact that $\dot{r}$ in Eq.~\eqref{dot-r} must vanish at the distance $r=r_0$ yields
\begin{equation}
b=\sqrt{\frac{\left(r_0^2+\eta^2\right)\left[\sqrt{r_0^2+\eta^2}+2\left(\frac{1}{w^2}-1\right)M\right]}{\sqrt{r_0^2+\eta^2}-2M}}~,  \label{b-r0}
\end{equation}
where the nonphysical solution has been ignored. By defining
\begin{equation}
h\equiv M/r_0~,~~~~~~~~~~\hat{\eta}\equiv \eta/M~,~~ \label{haQ}
\end{equation}
Eq.~\eqref{b-r0} is expanded as power series in the small parameter $h$
\begin{widetext}
\begin{equation}
b=r_0\left[1+\frac{h}{w^2}+\frac{1}{2}\left(\hat{\eta}^2+\frac{4}{w^2}-\frac{1}{w^4}\right)h^2+\frac{1-4w^2+8w^4}{2w^6}h^3\right]+\mathcal{O}(M^4)~, \label{b-r0-2}
\end{equation}
which conversely leads to
\begin{equation}
r_0=b\left[1-\frac{1}{w^2}\frac{M}{b}-\frac{1}{2}\left(\hat{\eta}^2+\frac{4}{w^2}-\frac{1}{w^4}\right)\left(\frac{M}{b}\right)^2
-\frac{8+\hat{\eta}^2}{2w^2}\left(\frac{M}{b}\right)^3\right]+\mathcal{O}(M^4)~.  \label{b-r0-3}
\end{equation}

On the other hand, according to Eqs.~\eqref{C1} - \eqref{dot-r}, the exact form of the bending angle can be expressed in the form
\begin{eqnarray}
\alpha=2\int_{r_0}^{+\infty}\left|\frac{d\varphi}{dr}\right|dr-\pi
=2\int_{r_0}^{+\infty}\!\frac{w\,b}{(r^2+\eta^2)\sqrt{1-\left(1-\frac{2M}{\sqrt{r^2+\eta^2}}\right)\left(1-w^2+\frac{w^2b^2}{r^2+\eta^2}\right)}}\,dr-\pi~,  \label{alpha-exact-1}
\end{eqnarray}
which can be rewritten in a convenient way by using a new variable $x\equiv r_0/r$ as
\begin{eqnarray}
\alpha=2\int_{0}^{1}\frac{w\,b}{r_0\left(1+\hat{\eta}^2h^2x^2\right)\sqrt{\hat{F}}}\,dx-\pi~, \label{alpha-exact-2}
\end{eqnarray}
with
\begin{equation}
\hat{F}=1-\left(1-\frac{2\,h\,x}{\sqrt{1+\hat{\eta}^2h^2x^2}}\right)
\left[1-w^2+\frac{w^2\sqrt{1+\hat{\eta}^2h^2}+2(1-w^2)h}{\sqrt{1+\hat{\eta}^2h^2}-2h}\frac{\left(1+\hat{\eta}^2h^2\right)x^2}{1+\hat{\eta}^2h^2x^2}\right]~.~~\label{H}
\end{equation}
Expanding the square root in Eq.~\eqref{alpha-exact-2} gives
\begin{eqnarray}
\frac{1}{\sqrt{\hat{F}}}=\frac{1}{w\sqrt{1-x^2}}\Bigg{\{}1\!+\!\left[1\!-\!\frac{1}{w^2\left(1\!+\!x\right)}\right]\!xh
+\frac{3\!-\!w^2\left(1\!+\!x\right)\left[2\!-\!w^2\!\left(3\!+\!\hat{\eta}^2\right)\left(1\!+\!x\right)\right]}{2w^4\left(1+x\right)^2}\,x^2h^2
+\frac{V_1\,x^2h^3}{2w^6\left(1\!+\!x\right)^3}\!\Bigg{\}}+\mathcal{O}(M^4)~,~~~~~~  \label{H2}
\end{eqnarray}
where
\begin{eqnarray}
V_1=\left(8\!+\!\hat{\eta}^2\right)\!w^4\!+\!\left[3w^2\!-\!5\!+\!\left(13\!+\!\hat{\eta}^2\right)\!w^4\!+\!5w^6\right]\!x
\!+\!w^2\!\left[3\!+\!\left(2\!-\!\hat{\eta}^2\right)w^2\!+\!15w^4\right]\!x^2\!-\!w^4\!\left(3\!-\!15w^2\!+\!\hat{\eta}^2\right)\!x^3\!+\!5w^6x^4~.~~~~~~ \label{V1}
\end{eqnarray}

The substitution of Eqs.~\eqref{b-r0-2} and \eqref{H2} into the power series expansion of Eq.~\eqref{alpha-exact-2} in $h$ yields
\begin{eqnarray}
&&\nn\alpha=2\left(1+\frac{1}{w^2}\right)h+\left[\frac{3\pi}{4}\left(1+\frac{4}{w^2}\right)-\frac{2}{w^2}\left(1+\frac{1}{w^2}\right)+\frac{\pi\,\hat{\eta}^2}{4}\right]h^2  \\
&&\hspace*{18pt}+\left[\frac{10}{3}+\frac{26}{w^2}+\frac{9}{w^4}+\frac{7}{3w^6}-\frac{3\pi}{2w^2}\left(1+\frac{4}{w^2}\right)
-\left(\frac{1}{3}+\frac{\pi-2}{2w^2}\right)\hat{\eta}^2\right]h^3+\mathcal{O}(M^4)~.  \label{alpha-3PM}
\end{eqnarray}
By substituting Eqs.~\eqref{haQ} and \eqref{b-r0-3} into Eq.~\eqref{alpha-3PM}, the analytical form of the gravitational deflection angle up to the 3PM order of a relativistic massive particle $(0.05\lesssim w<1)$ in the black-bounce-Schwarzschild spacetime is eventually achieved as follows:
\begin{eqnarray}
&&\alpha=2\!\left(\!1+\frac{1}{w^2}\!\right)\!\frac{M}{b}+\frac{3\pi}{4}\!\left(\!1+\frac{4}{w^2}\!\right)\!\frac{M^2}{b^2}+\frac{\pi}{4}\frac{\eta^2}{b^2}
+\frac{2}{3}\!\left(\!5+\frac{45}{w^2}+\frac{15}{w^4}-\frac{1}{w^6}\!\right)\!\frac{M^3}{b^3}
+\frac{2}{3}\!\left(\!1+\frac{3}{w^2}\!\right)\!\frac{M\eta^2}{b^3}+\mathcal{O}(M^4)~,~~~~~~~~  \label{alpha-3PM-Final}
\end{eqnarray}
which is equivalent to
\begin{equation}
\alpha=C_1\frac{M}{b}+C_2\frac{M^2}{b^2}+C_3\frac{M^3}{b^3}+\mathcal{O}(M^4)~,  \label{alpha-3PM-Final-2}
\end{equation}
with
\begin{equation}
C_1=2\left(1+\frac{1}{w^2}\right)~,~~~~~~
C_2=\frac{\pi}{4}\left[3\left(1+\frac{4}{w^2}\right)+\hat{\eta}^2\right]~,~~~~~~
C_3=\frac{2}{3}\left(5+\frac{45}{w^2}+\frac{15}{w^4}-\frac{1}{w^6}\right)+\frac{2}{3}\left(1+\frac{3}{w^2}\right)\!\hat{\eta}^2~.  \label{C}
\end{equation}

Since the deflection angle diverges and the weak-field and small-angle approximation breaks down in the limit $w\rightarrow 0$, it should be emphasized that Eq.\,\eqref{alpha-3PM-Final} applies only to test particles which are relativistic. For example, if $w=0.1$ and $M/b=1.0\times10^{-5}$ (a typical weak field) are assumed, then the deflection
is about $0.12^{\,\circ}$, where the small-angle approximation still holds. Additionally, we see that both the second- and third-order contributions from the bounce on the right-hand side of Eq.\,\eqref{alpha-3PM-Final} to the gravitational deflection of massive particles are positive. When $w=1$, Eq.\,\eqref{alpha-3PM-Final} is simplified to the black-bounce-Schwarzschild deflection angle of light~\cite{CX2021,ZX2022,Ali2020a,GL2023}
\begin{equation}
\left. \alpha\right|_{w=1}=\frac{4M}{b}+\frac{15\pi}{4}\frac{M^2}{b^2}+\frac{\pi}{4}\frac{\eta^2}{b^2}+\frac{128}{3}\frac{M^3}{b^3}+\,\frac{8}{3}\frac{M\eta^2}{b^3}
+\mathcal{O}(M^4)~. \label{alpha-3PM-Light}
\end{equation}
Furthermore, Eq.\,\eqref{alpha-3PM-Final} can be reduced to the third-order Schwarzschild deflection angle of massive particles~\cite{AR2002,AR2003,LZLH2019,HL2022}, if the bounce parameter of the spacetime disappears.

\section{Observable lensing properties} \label{sect4}
We now consider the weak-field lensing characteristics of the primary and secondary images of a point-like massive-particle source (for more details see~\cite{PNHZ2014,LYJ2016,BT2017,PJ2019,Frost2023}), for which the standard perturbation theory analysis is adopted to solve the Virbhadra-Ellis lens equation~\cite{VE2000}
\begin{equation}
\tan\mathcal{B}=\tan\vartheta-D\left[\tan\vartheta+\tan(\alpha-\vartheta)\right]~,  \label{LE}  \vspace*{7pt}
\end{equation}
which applies to both the weak- and strong-field motions of test particles and is obtained from Fig.~\ref{Figure1} by defining $D=d_{LS}/d_{S}$.

To do this, we first define three scaled quantities for convenience by~\cite{KP2005,KP2006a,KP2006b}
\begin{equation}
\beta\equiv\frac{\mathcal{B}}{\vartheta_E}~,~~~~~~~~~~~~\theta\equiv\frac{\vartheta}{\vartheta_E}~,~~~~~~~~~~~~
\varepsilon\equiv\frac{\vartheta_{\bullet}}{\vartheta_E}=\frac{\vartheta_E}{4D}~,   \label{variables}
\end{equation}
where $\vartheta_E \,(=\!\sqrt{4DM/d_L}\,)$ and $\varepsilon$ denote the weak-deflection angular Einstein radius of light and a new expansion parameter for analyzing the lensing observables, respectively. Moreover, $\vartheta_{\bullet}$ is the angle subtended by the special gravitational radius $M_\bullet \,(\equiv GM/c^2=M)$~\cite{KP2005}, with $\vartheta_{\bullet}=\arctan\left(M_\bullet/d_L\right)$. Note that here we follow the treatment of~\cite{HL2022} to regard $\vartheta_E$ in Eq.~\eqref{variables} as the natural scale, which can ensure that the velocity effects on the angular image position are absorbed completely by the scaled quantity $\theta$. Note also that Eq.~\eqref{LE} will be simplified to the small angles lens equation $\vartheta=\mathcal{B}+\hat{\alpha}$~\cite{SEF1992}, if we define the reduced deflection angle by $\hat{\alpha}\equiv D \alpha$ and neglect the third- and higher-order contributions in $\varepsilon$ to the angular quantities.

Then, the perturbation analysis indicates that the scaled angular image position can be expressed as a power series in the small parameter $\varepsilon$
\begin{equation}
\theta=\theta_0+\theta_1\varepsilon+\theta_2\varepsilon^2+\mathcal{O}(\varepsilon^3)~.   \label{theta}
\end{equation}
Here, $\theta_0$ is the weak-deflection image position which is positive. $\theta_1$ and $\theta_2$ represent the undetermined coefficients of the first- and second-order corrections to the image position respectively. The substitution of $b=d_L\sin\vartheta$ and Eqs.~\eqref{alpha-3PM-Final-2}, \eqref{variables}, and \eqref{theta} into Eq.~\eqref{LE} thus gives
\begin{eqnarray}
\nn&&0=D\left(4\beta-4\theta_0+\frac{C_1}{\theta_0}\right)\varepsilon+\frac{D}{\theta_0^2}\!\left[C_2-\left(C_1+4\theta_0^2\right)\theta_1\right]\varepsilon^2
+\,\frac{D}{3\theta_0^3}\Big{[}C_1^3+3C_3-12DC_1^2\theta_0^2+C_1\!\left(56D^2\theta_0^4+3\theta_1^2-3\theta_0\theta_2\right)    \\
&&\hspace*{18pt}+\,64D^2\theta_0^3\left(\beta^3-\theta_0^3\right)-6C_2\theta_1-12\theta_0^3\theta_2\Big{]}\varepsilon^3+\mathcal{O}(\varepsilon^4)~.   \label{LE-2}
\end{eqnarray}

\subsection{Image positions}  \label{IP}
We then focus on the explicit form of the solution of the lens equation. By requiring each of the terms on the right-hand side of Eq.~\eqref{LE-2} to vanish, we obtain
\begin{eqnarray}
&&\theta_0=\frac{1}{2}\left(\beta+\xi\right)~,    \label{theta0} \\
&&\theta_1=\frac{\pi\!\left[3\left(4+w^2\right)+w^2\hat{\eta}^2\right]}{8\left(1+w^2+2w^2\theta_0^2\right)}~,   \label{theta1}  \\
&&\nn\theta_2=\frac{1}{24\,w^{10}\,\theta_0\left(2+\frac{2}{w^2}+4\theta_0^2\right)^3}\bigg{\{}256\left(1-D^2\right)\left(1+w^2\right)^5
-3w^4\left(1+w^2\right)\left[3\pi\left(4+w^2\right)+\pi w^2\hat{\eta}^2\right]^2   \\
&&\nn\hspace*{23pt}-\,64\left(1\!+\!w^2\right)^2\left[1\!-\!5w^2\left(3\!+\!9w^2\!+\!w^4\right)-w^4\left(3+w^2\right)\!\hat{\eta}^2\right]
+4w^2V_2\,\theta_0^2-256w^4V_3\,\theta_0^4-2048\,w^6   \\
&&\hspace*{23pt}\times\left(1+w^2\right)^2D\left(3-4D\right)\theta_0^6+2048w^8\left(1+w^2\right)D^2\theta_0^8\bigg{\}}~,~~~~~   \label{theta2}
\end{eqnarray}
with
\begin{eqnarray}
&&\xi=\sqrt{\beta^2+2\left(1+\frac{1}{w^2}\right)}~, \label{xi}  \\
&&\nn V_2=128\left(1+w^2\right)^4(1-D)\,(2-D)-\,64\left(1+w^2\right)\left[1-5w^2\left(3+9w^2+w^4\right)-w^4\left(3+w^2\right)\hat{\eta}^2\right] \\
&&\hspace*{23pt}-3w^4\!\left[3\pi\left(4+w^2\right)+\pi w^2\hat{\eta}^2\right]^2~,  \label{V2} \\
&&V_3=\,1-15w^2-45w^4-5w^6-2\left(1+w^2\right)^3\left[2-D\left(12-11D\right)\right]-w^4\!\left(3+w^2\right)\hat{\eta}^2~.~~~~~~ \label{V3}
\end{eqnarray}
Since an angular image position has been assumed to be always positive (see Section~\ref{sect2} for more details), the scaled positions of the positive-parity primary image and the negative-parity secondary image of the particle source can be rewritten respectively in terms of the scaled source position
\begin{equation}
\theta^{\pm}=\theta_0^{\pm}+\theta_1^{\pm}\varepsilon+\theta_2^{\pm}\varepsilon^2+\mathcal{O}(\varepsilon^3)~,   \label{theta-PN}
\end{equation}
where
\begin{eqnarray}
&&\theta_0^{\pm}=\frac{1}{2}\left(\xi\pm\left|\beta\right|\right)~,    \label{theta0-PN}  \\
&&\theta_1^{\pm}=\frac{\pi\!\left[3\left(4+w^2\right)+w^2\hat{\eta}^2\right]}{16\left(1+w^2\right)}\!\left(1\mp\frac{\left|\beta\right|}{\xi}\right)~,~~~~~~ \label{theta1-PN} \\
&&\theta_2^{\pm}=\frac{V_4}{w^6\,\xi\left(\xi\pm\left|\beta\right|\right)^2}
-\frac{4\left(1+w^2\right)^2 D}{w^4\,\xi}\!\left[1-D-\frac{w^2D\left(\xi\pm\left|\beta\right|\right)^2}{12\left(1+w^2\right)}\right]~,~~~~~~~  \label{theta2-PN}
\end{eqnarray}
and
\begin{eqnarray}
V_4=2\left(1\!+\!3w^2\right)\!\left(1\!+\!6w^2\!+\!w^4\right)\!-\!\frac{8\left(1\!+\!w^2\right)^3\!D^2}{3}\!+\!\frac{2w^4\!\left(3\!+\!w^2\right)\!\hat{\eta}^2}{3}
\!-\!\frac{\pi^2w^4\!\left[3\!\left(4\!+\!w^2\right)\!+\!w^2\hat{\eta}^2\right]^2}{128\left(1+w^2\right)}\!\bigg{(}1\mp\frac{\left|\beta\right|}{\xi}\bigg{)}\!\bigg{(}3\pm\frac{\left|\beta\right|}{\xi}\bigg{)}
~.~~~~~~ \label{V4}
\end{eqnarray}
\end{widetext}
It is interesting to find that Eqs.~\eqref{theta0} - \eqref{theta2} match well with the result given in~\cite{ZX2022} for the case of light, and are in agreement with the result of~\cite{AKP2011} when $w=1$ and $\hat{\eta}=0$. Moreover, Eqs.\,(24) - (26) in~\cite{CX2021} can be recovered from Eqs.~\eqref{theta0-PN} - \eqref{theta2-PN} for the case of $w=1$. When the bounce parameter is omitted ($\hat{\eta}=0$), Eqs.~\eqref{theta0-PN} - \eqref{theta2-PN} are consistent with Eqs.\,(42) - (44) of~\cite{HL2022}, and Eqs.~\eqref{theta0-PN} - \eqref{theta1-PN} are in accordance with Eq.\,(71) in~\cite{PJ2019}. Additionally, Eq.~\eqref{theta-PN} means that the coefficients of the first- and second-order contributions to the positions of the primary and secondary images of the particle source have the bounce dependence. Thus, we may measure or constrain the bounce parameter of the spacetime conversely through the detections of the first- and second-order positional coefficients, when an astrophysical black hole is properly modeled as a black-bounce-Schwarzschild black hole.
\begin{figure*}
\centering
\begin{minipage}[b]{13cm}
\includegraphics[width=13cm]{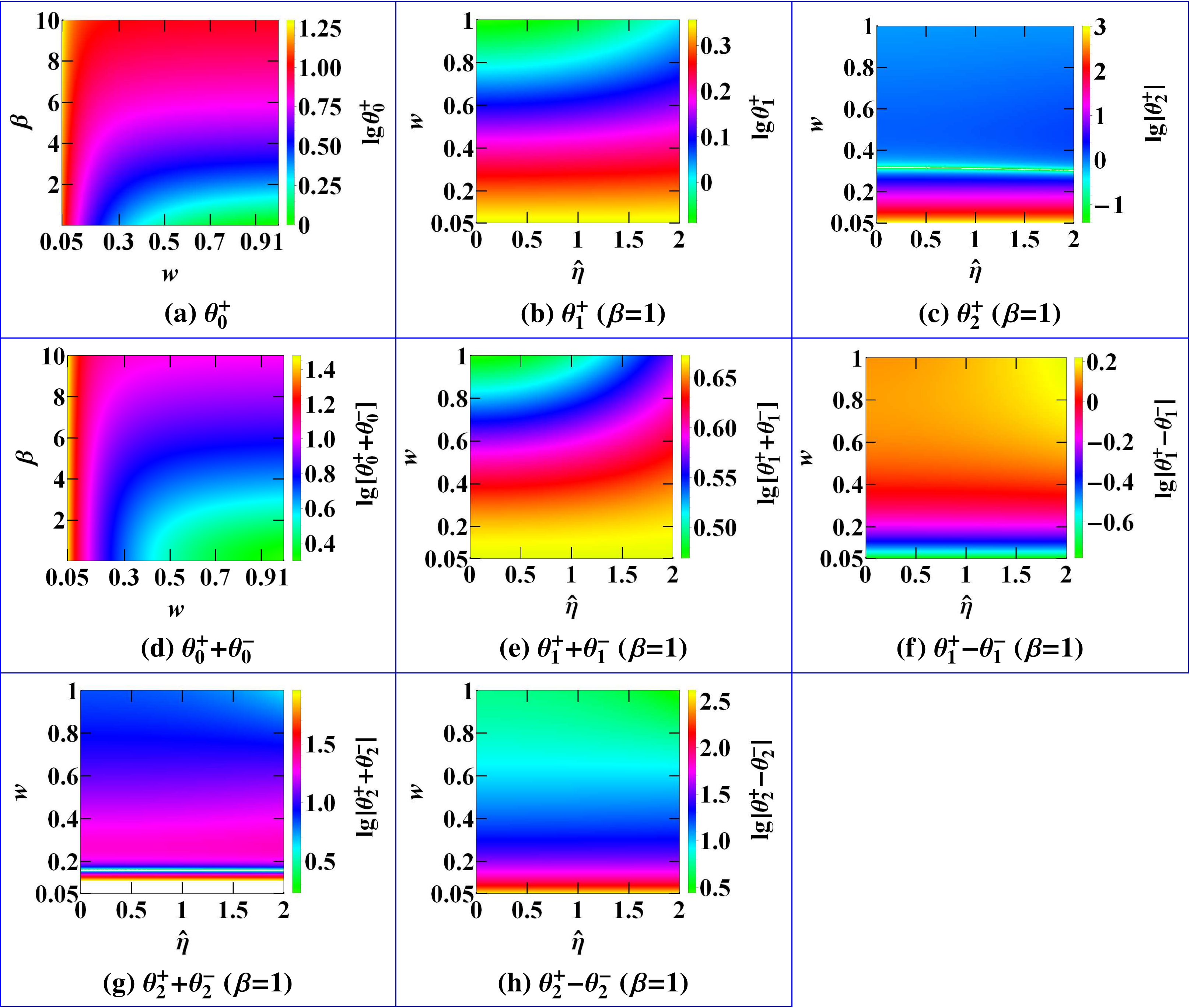}
\end{minipage}
\caption{The coefficients of the scaled angular image position and the positional sum and difference relations plotted as the functions of $w$ and $\beta$ or $\hat{\eta}$ and $w$ in color-indexed form for $D=0.5$. The first- and second-order quantities among them are shown for the case of $\beta=1$. We adopt the notation in the grid to show this case, and a similar treatment applies to all of the following figures. Here and hereafter, the initial velocity of a relativistic test particle is assumed to have a mentioned rough range $0.05\lesssim w\leq1$ for the convenience of discussion. Additionally, $\beta\in[0.01,~10]$ and $\hat{\eta}\in(0,~2)$ are assumed as an example of our black-bounce-Schwarzschild lensing scenario of massive particles. Moreover, as explained in Section VI of~\cite{HL2022}, a white region in a color-indexed figure denotes the value domain where the value of a lensing quantity is too large or too small to be displayed properly, which does not affect our analysis. Notice that we focus on the absolute values of the lensing quantities, and that the notation $w=1$ in the grid indicates the null case of a lensing observable hereafter. } \label{Figure2}
\end{figure*}

The relations among the coefficients of the scaled image positions ($\theta^+$ and $\theta^-$) are also indicated according to Eq.~\eqref{theta-PN}, and they are
\begin{eqnarray}
&&\theta_0^{+}\theta_0^{-}=\frac{1}{2}\left(1+\frac{1}{w^2}\right)~, \label{NewR-1}  \\
&&\theta_0^{+}+\theta_0^{-}=\xi~,~~~~ \label{NewR-2}  \\
&&\theta_0^{+}-\theta_0^{-}=\left|\beta\right|~, \label{NewR-3} \\
&&\theta_1^{+}+\theta_1^{-}=\frac{\pi\!\left[3(4+w^2)+w^2\hat{\eta}^2\right]}{8(1+w^2)}~, \label{NewR-4}  \\
&&\theta_1^{+}-\theta_1^{-}=-\frac{\pi\!\left[3(4+w^2)+w^2\hat{\eta}^2\right]}{8(1+w^2)}\frac{\left|\beta\right|}{\xi}~,~~~~~~~ \label{NewR-5} \\
&&\theta_2^{+}+\theta_2^{-}=\frac{V_5+V_6\,w^4\hat{\eta}^2}{192w^6\left(1\!+\!w^2\right)^3\!\xi^3}~,  \label{NewR-6}   \\
&&\theta_2^{+}-\theta_2^{-}=-\frac{V_7\left|\beta\right|}{96w^2\left(1+w^2\right)^3}~,    \label{NewR-7}
\end{eqnarray}
where
\begin{widetext}
\begin{eqnarray}
&&\nn V_5=768+1152\,w^2\left(8+\beta^2\right)
-\left(1+w^2\right)^4w^2\xi^2\!\left[1536\,D\left(1+w^2\right)-256\,D^2\left(5+5\,w^2-w^2\beta^2\right)\right]-27\,\pi^2w^4  \hspace*{12pt}  \\
&&\nn\hspace*{21pt}\times\left(4+w^2\right)^2\!\left[3+6w^2\!\left(1+\beta^2\right)+w^4\!\left(3+6\beta^2+2\beta^4\right)\right]+384w^4\big{[}2\left(49+88w^2+75w^4+28w^6+3w^8\right)   \\
&&\hspace*{21pt}+\,3\left(11+38w^2+50w^4+25w^6+3w^8\right)\beta^2+\left(1+10w^2+28w^4+22w^6+3w^8\right)\beta^4\big{]}~, \label{V5}   \\
&&\nn V_6=4w^2\beta^2\!\left[3(1\!+\!w^2)\!+\!w^2\beta^2\right]\!\!\left[96\!+\!(32\!-\!9\pi^2)w^2(4\!+\!w^2)\right]
\!-\!3\pi^2w^4\!\left[3(1\!+\!w^2)^2\!+\!6w^2(1\!+\!w^2)\beta^2\!+\!2\,w^4\beta^4\right]\!\hat{\eta}^2  \\
&&\hspace*{21pt}+\,2\left(1+w^2\right)^2\left[384+\left(128-27\pi^2\right)w^2\left(4+w^2\right)\right]~, \label{V6} \\
&&\nn V_7=3\!\left[64+640w^2+16\left(112-9\pi^2\right)\!w^4+8\left(176-9\pi^2\right)\!w^6+3\left(64-3\pi^2\right)\!w^8-128D^2\!\left(1+w^2\right)^4\right]+2w^4 \\
&& \hspace*{21pt}\times\!\left[96+\left(32-9\pi^2\right)w^2\left(4+w^2\right)\right]\!\hat{\eta}^2-3\pi^2w^8\hat{\eta}^4~.  \label{V7}
\end{eqnarray}
\end{widetext}
It should be noted that the first- and second-order position relations given in Eqs.~\eqref{NewR-4} - \eqref{NewR-7} are dependent on the bounce parameter, in contrast to the case of the zeroth-order components shown in Eqs.~\eqref{NewR-1} - \eqref{NewR-3}. Moreover, similar to the lensing scenario of timelike signals in Kerr-Newman spacetime~\cite{HL2022}, those position relations (except the zeroth-order positional difference) depend on the initial velocity of the particle, which leaves us the opportunity to probe the characteristics of the particle source in turn. Figure~\ref{Figure2} shows the zeroth-, first-, and second-order positional coefficients of the primary image of the particle source, accompanied by the positional sum and difference relations, as the bivariate functions of the scaled source position (or the scaled bounce parameter) and the initial velocity of the particle.

\subsection{Magnifications} \label{Mag}
Next we go on to the magnifications of the lensed images of the point source of massive particles. The magnification of an image is defined by the ratio of the flux of the image to that of the unlensed source, and it is equivalent to the ratio between the solid angles of the image and the source and thus reads~\cite{VE2000,ERT2002,PNHZ2014,PJ2019}
\begin{eqnarray}
\mu(\vartheta)=\frac{\sin\vartheta}{\sin\mathcal{B}}\frac{d\vartheta}{d\mathcal{B}}~,  \label{mu-Define}
\end{eqnarray}
which yields on the basis of Eqs.~\eqref{alpha-3PM-Final-2} and \eqref{LE}
\begin{equation}
\mu=\mu_0+\mu_1\varepsilon+\mu_2\varepsilon^2+\mathcal{O}(\varepsilon^3)~,   \label{mu-SeriesE}
\end{equation}
with the coefficients of the zeroth-, first-, and second-order contributions to the magnification being respectively given by
\begin{eqnarray}
&&\mu_0=\frac{4\theta_0^4}{4\theta_0^4-\left(1+\frac{1}{w^2}\right)^2}~,  \label{mu0}  \\
&&\mu_1=-\frac{\pi w^4\!\left[3(4+w^2)+w^2\hat{\eta}^2\right]\!\theta_0^3}{2(1+w^2+2w^2\theta_0^2)^3}~,~~~~~~~  \label{mu1}
\end{eqnarray}
\begin{widetext}
\begin{equation}
\mu_2=-\frac{\theta_0^2\,\Big{\{}16\left(1+w^2\right)^6D^2+V_8\,\theta_0^2-w^4V_9\,\theta_0^4+128w^6V_{10}\,\theta_0^6
+256w^8\left(1+w^2\right)^2D^2\theta_0^8\Big{\}}}{6w^{12}\left(1+\frac{1}{w^2}-2\theta_0^2\right)\left(1+\frac{1}{w^2}+2\theta_0^2\right)^5}~.~~~ \label{mu2}
\end{equation}
Here, the parameters $V_8$, $V_9$, and $V_{10}$ are defined by
\begin{eqnarray}
&&V_8=32w^2\!\left(1+w^2\right)^2\left[1-5w^2\!\left(3+9w^2+w^4\right)-w^4\!\left(3+w^2\right)\hat{\eta}^2\right]-64w^2\!\left(1+w^2\right)^5\left(2+6D-9D^2\right)~,~~~~~~ \label{V8} \\
&&V_9=128\left(1+w^2\right)^4\left[4+D\left(12-17D\right)\right]-128\left(1+w^2\right)\left[1-5w^2\left(3+9w^2+w^4\right)-w^4\left(3+w^2\right)\hat{\eta}^2\right]   \\
&&\hspace*{23pt}-\,9w^4\!\left[3\pi\left(4+w^2\right)+\pi w^2\hat{\eta}^2\right]^2~,~~ \label{V9}  \\
&&V_{10}=1-5w^2\left(3+9w^2+w^4\right)-w^4\left(3+w^2\right)\hat{\eta}^2-2\left(1+w^2\right)^3\left(2+6D-9D^2\right)~.~~ \label{V10}
\end{eqnarray}

Since the magnification sign of an image indicates its parity, we can easily check that the magnifications ($\mu^+$ and $\mu^-$) of the positive-parity primary image and the negative-parity secondary image are positive and negative, respectively. Expressed in terms of the source position $\beta$, the coefficients of the zeroth-, first-, and second-order contributions to $\mu^+$ (and $\mu^-$) are then obtained as follows:
\begin{eqnarray}
&&\mu_0^{\pm}=\frac{1}{2}\pm\frac{1}{2\left|\beta\right|\xi}\!\left(1+\frac{1}{w^2}+\beta^2\right)~,  \label{mu0-beta}  \\
&&\mu_1^{\pm}=-\frac{\pi\!\left[3\left(4+w^2\right)+w^2\hat{\eta}^2\right]}{16w^2\xi^3}~,~~~~  \label{mu1-beta}  \\
&&\mu_2^{\pm}=\frac{\left\{\left(1+\frac{1}{w^2}\right)^3+\beta^2\left[3\left(1+\frac{1}{w^2}\right)+2\beta^2\right]^2
\pm\left[3\left(1+\frac{1}{w^2}\right)^2+8\left(1+\frac{1}{w^2}\right)\beta^2+4\beta^4\right]\xi\left|\beta\right|\right\}V_{11}}
{3\,\xi^5\left(\beta^2\pm\xi\left|\beta\right|\right)\left(\xi\pm\left|\beta\right|\right)^5}~,~~~~ \label{mu2-beta}
\end{eqnarray}
where
\begin{eqnarray}
&&\nn V_{11}=-\frac{8\,\xi^2}{w^6}\!\left[3\left(1+9w^2+19w^4+3w^6\right)+12\left(1+w^2\right)^3D
+w^4\left(3+w^2\right)\hat{\eta}^2-\,2\left(9+9w^2+w^2\beta^2\right)\left(1+w^2\right)^2D^2\right]  \\
&&\hspace*{29pt}+\,\frac{9\pi^2}{16}\!\left(3+\frac{12}{w^2}+\hat{\eta}^2\right)^2~.~ \label{V11}
\end{eqnarray}
\end{widetext}
Eqs.~\eqref{mu0-beta} - \eqref{mu2-beta} further yield the following magnification relations
\begin{eqnarray}
&&\mu_0^{+}+\mu_0^{-}=1~, \label{mu0-PPN}  \\
&&\mu_0^{+}-\mu_0^{-}=\frac{1}{\xi\left|\beta\right|}\left(1+\frac{1}{w^2}+\beta^2\right)~,  \label{mu0-PMN}  \\
&&\mu_1^{+}+\mu_1^{-}=2\mu_1^{+}~,~~~~~ \label{mu1-PPN} \\
&&\mu_2^{+}-\mu_2^{-}=2\mu_2^{+}~. ~~~~~~  \label{mu2-PNN}
\end{eqnarray}

With respect to Eqs.~\eqref{mu0} - \eqref{mu2-PNN}, three points should be emphasized. First, we find that Eqs.~\eqref{mu0} - \eqref{mu2-beta} can reduce to the null result in black-bounce-Schwarzschild geometry~\cite{CX2021,ZX2022} in the limit $w\rightarrow1$, and that Eqs.~\eqref{mu0} - \eqref{mu2} are consistent with the null result in Schwarzschild spacetime~\cite{KP2005} when $w=1$ and no bounce effect is considered $(\eta=0)$. Moreover, it is also found that Eqs.~\eqref{mu0} - \eqref{mu2-PNN} are consistent with the results presented in~\cite{HL2022} for the Schwarzschild lensing scenario of massive particles, when removing the contribution from the bounce parameter. Second, the universal magnification relations $\mu_1^+=\mu_1^-$ and $\mu_2^+=-\mu_2^-$ for a static and spherically symmetric spacetime~\cite{KP2005,KP2006a,KP2006b,WP2007},
which lead to Eqs.~\eqref{mu1-PPN} - \eqref{mu2-PNN}, apply not only to the case of electromagnetic waves but also to the case of massive particles.
Third, we notice that all of the zeroth-, first-, and second-order contributions to $\mu^+$ or $\mu^-$ depend on the initial velocity of the massive particle. It is also true for the zeroth-order difference, first-order sum and second-order difference relations of the magnification coefficients. Additionally, the dependence on the bounce parameter appears only for the first- and second-order magnifications, as well as for the first-order sum and second-order difference relations of the magnifications.

\begin{figure*}
\centering
\begin{minipage}[b]{13cm}
\includegraphics[width=13cm]{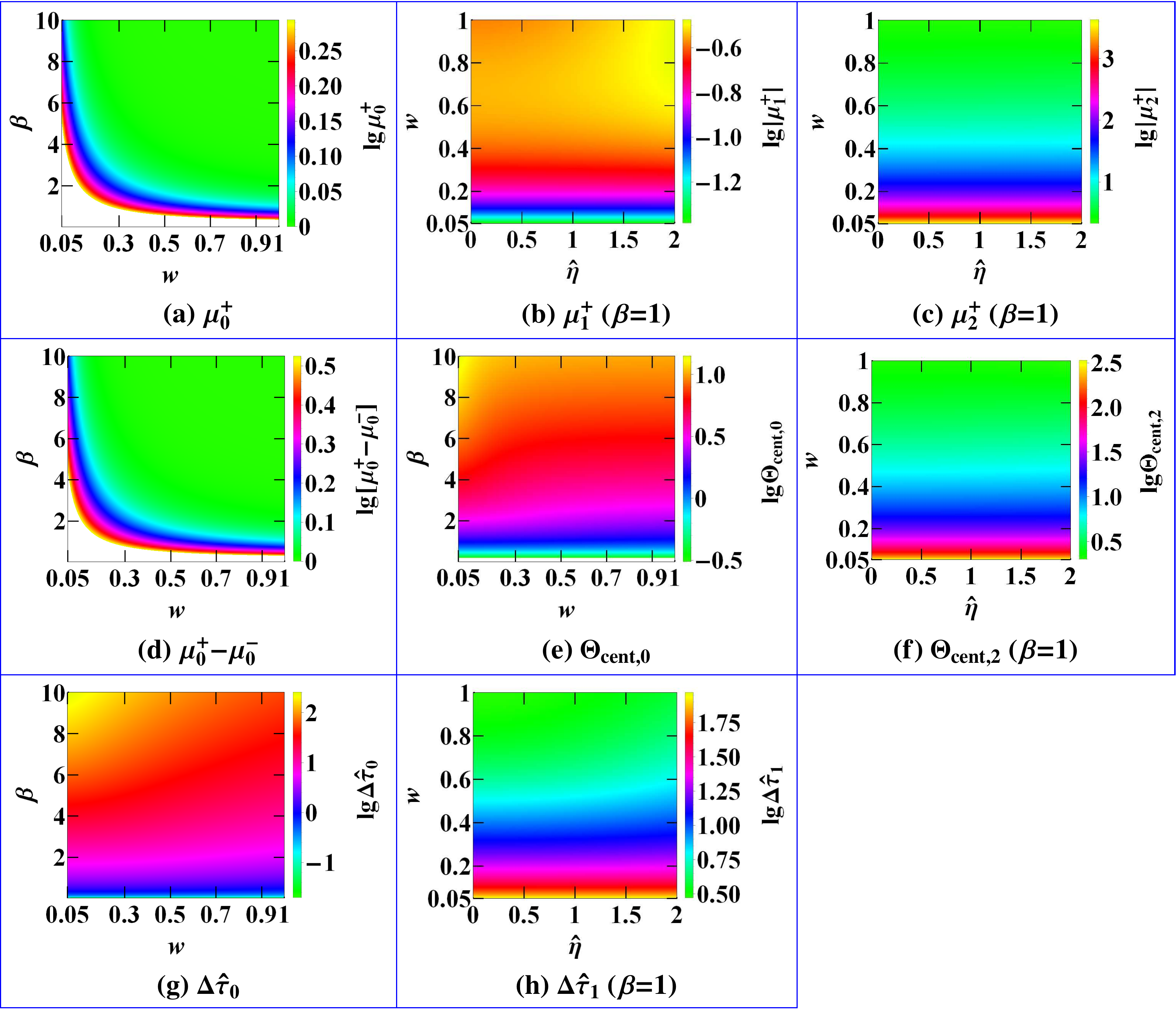}
\end{minipage}
\caption{The coefficients of the image magnification $\mu^+$, the scaled centroid $\Theta_{\text{cent}}$, and the scaled differential time delay $\Delta\hat{\tau}$, together with the zeroth-order magnification difference, plotted as the color-indexed functions of $w$ and $\beta$ or $\hat{\eta}$ and $w$ for $D=0.5$. } \label{Figure3}
\end{figure*}

Figure\,\ref{Figure3} shows the coefficients of the zeroth-, first-, and second-order contributions to $\mu^+$, along with the zeroth-order difference relation given in Eq.~\eqref{mu0-PMN}, as the bivariate functions of the scaled source position (or the scaled bounce parameter) and the initial velocity of the particle. \\

\subsection{Total magnification and centroid}  \label{TMagCent}
If the two images are difficult to be resolved, then the sum of the absolute values of their magnifications and the magnification-weighted centroid position are the useful observables. The definition of the measurable total magnification reads
\begin{equation}
\mu_{\text{tot}}\equiv|\mu^{+}|+|\mu^{-}|=\mu^{+}-\mu^{-}~,  \label{mu-tot}
\end{equation}
which is always larger than 1 and is expressed as up to the second order in $\varepsilon$
\begin{widetext}
\begin{eqnarray}
\mu_{\text{tot}}=\frac{1}{\xi\left|\beta\right|}\!\left(\!1+\frac{1}{w^2}+\beta^2\!\right)
+\frac{V_{12}\,\varepsilon^2}{12\,w^8\,\xi^5\left|\beta\right|}+\mathcal{O}(\varepsilon^3)~,~~~~~~~ \label{mu-tot-Series}
\end{eqnarray}
with
\begin{eqnarray}
&&\nn V_{12}=-8w^2\xi^2\!\left[12D\left(1+w^2\right)^3+3\left(1+3w^2\right)\left(1+6w^2+w^4\right)-2D^2\left(1+w^2\right)^2\left(9+9w^2+w^2\beta^2\right)
+w^4\left(3+w^2\right)\hat{\eta}^2\right]   \\
&& \hspace*{29pt}+\,\frac{9\,\pi^2w^4}{16}\!\left(12+3w^2+w^2\hat{\eta}^2\right)^2~.~ \label{V12}
\end{eqnarray}

The scaled magnification-weighted centroid position of the images is defined by
\begin{equation}
\Theta_{\text{cent}}=\frac{\theta^{+}|\mu^{+}|-\theta^{-}|\mu^{-}|}{|\mu^{+}|+|\mu^{-}|}~,  \label{centroid}
\end{equation}
and its explicit form can be obtained as in terms of $\beta$

\begin{equation}
\Theta_{\text{cent}}=\Theta_{\text{cent,0}}+\Theta_{\text{cent,2}}\,\varepsilon^2+\mathcal{O}(\varepsilon^3)~, \label{centroid-Series-2}
\end{equation}
where
\begin{eqnarray}
&&\Theta_{\text{cent,0}}=\frac{\left[3\left(1+w^2\right)+2w^2\beta^2\right]\left|\beta\right|}{2\left(1+w^2+w^2\beta^2\right)}~,  \label{centroid-Series-beta-1}  \\
&&\Theta_{\text{cent,2}}=\frac{V_{13}V_{14}}{3w^{14}\xi^2\left(\xi+\left|\beta\right|\right)^3\!\left[4\left(1+\frac{1}{w^2}\right)^2+\left(\xi+\left|\beta\right|\right)^4\right]^2}~, \label{centroid-Series-beta-3}
\end{eqnarray}
with
\begin{eqnarray}
&&\nn V_{13}=\left[\left(1+w^2\right)^3+12w^2\left(1+w^2\right)^2\beta^2+20\,w^4\left(1+w^2\right)\beta^4+8\,w^6\beta^6\right]\xi\left|\beta\right|
+7\left(1+w^2\right)^3\!\beta^2+28w^2\left(1+w^2\right)^2\beta^4    \\
&&\hspace*{28.7pt}+\,28w^4\left(1+w^2\right)\beta^6+8w^6\beta^8~,~ \label{V13}  \\
&&\nn V_{14}=w^4\!\left[768\!+\!2\left(128\!-\!27\pi^2\right)w^2\left(4\!+\!w^2\right)\!-\!9\pi^2w^4\hat{\eta}^2\right]\!\hat{\eta}^2
\!+\!768\left(1\!+\!10w^2\right)\!+\!48\left(448\!-\!27\pi^2\right)w^4\!+\!24\left(704\!-\!27\pi^2\right)w^6   \\
\nn&&\hspace*{26pt}+\,9\left(256-9\pi^2\right)w^8+128w^2\!\left[3\left(1+3w^2\right)\left(1+6w^2+w^4\right)+w^4\left(3+w^2\right)\hat{\eta}^2+2\left(1+w^2\right)^3\!D\left(17D-12\right)\right]\!\beta^2   \\
&&\hspace*{26pt}-\,256w^4\left(1+w^2\right)^2\left(6-13D\right)D\beta^4+512w^6\left(1+w^2\right)D^2\beta^6-512\left(1+w^2\right)^4D^2~.~    \label{V14}
\end{eqnarray}

With regard to Eqs.~\eqref{mu-tot-Series} and \eqref{centroid-Series-2}, it should be noted that they agree with the results of the Schwarzschild lensing of massive particles~\cite{HL2022} when removing the bounce effect. We also note that Eq.~\eqref{centroid-Series-2} for the case of $w=1$ can recover the following null result of the black-bounce-Schwarzschild lensing scenario~\cite{CX2021,ZX2022}
\begin{eqnarray}
\nn&&\Theta_{\text{cent}}=\frac{\left(3+\beta^2\right)\left|\beta\right|}{2+\beta^2}-\frac{\left|\beta\right|}{384\left(4+\beta^2\right)\left(2+\beta^2\right)^2}
\Big{\{}2025\pi^2\!-\!6144\left(2\!-\!D\beta^2\right)\left(4\!+\!\beta^2\right)\!+\!1024D^2\left(8\!-\!34\beta^2\!-\!13\beta^4\!-\!\beta^6\right)  \\
&&\hspace*{33pt}+\left[270\pi^2-512(4+\beta^2)\right]\hat{\eta}^2+9\pi^2\hat{\eta}^4\Big{\}}\,\varepsilon^2+\mathcal{O}(\varepsilon^3)~,~~~~~~~~  \label{centroid-w}
\end{eqnarray}
and that Eqs.~\eqref{mu-tot-Series} and \eqref{centroid-Series-2} are in accord with the null results of the Schwarzschild lensing scenario~\cite{KP2005} in the limit $w\rightarrow1$ and $\eta\rightarrow0$.
Additionally, Figure\,\ref{Figure3} shows the coefficients of the zeroth- and second-order contributions to the magnification centroid.

\subsection{Time delay}
We now consider the travel time of a massive particle propagating from the source to the observer, and then utilize a classical approach~\cite{We1972,KP2005} to calculate the difference between the gravitational time delays of the images of the particle source. Not to be forgotten, the differential time delay between two lensed images of a light source acts as a historically important lensing observable for its fruitful astronomical applications (see, e.g.,~\cite{Refsdal1964,Mao1992,FLBPZ2017}).

The combination of Eqs.~\eqref{C1} - \eqref{dot-r} yields
\begin{eqnarray}
\left|\frac{dt}{dr}\right|=\frac{1}
{b\left(1-\frac{2M}{\sqrt{r^2+\eta^2}}\right)\sqrt{\frac{1}{b^2}
-\frac{1}{r^2+\eta^2}\left(1-\frac{2M}{\sqrt{r^2+\eta^2}}\right)\left[w^2+\frac{(1-w^2)(r^2+\eta^2)}{b^2}\right]}}~, \label{dt-dr}
\end{eqnarray}
which gives immediately the propagating time of a massive particle travelling from the point of closest approach to a finite point (with a radial coordinate $R$) of the perturbed ray
\begin{eqnarray}
T(R)=\!\int_{r_0}^R\left|\frac{dt}{dr}\right|dr=r_0\!\!\int_{\frac{r_0}{R}}^{1}\!\frac{1}{x^2\!\left(1\!-\!\frac{2hx}{\sqrt{1+\hat{\eta}^2h^2x^2}}\right)\!
\sqrt{1\!-\!\frac{1}{1+\hat{\eta}^2h^2x^2}\!\left(\!1\!-\!\frac{2hx}{\sqrt{1+\hat{\eta}^2h^2x^2}}\!\right)
\!\!\left[\frac{w^2b^2}{r_0^2}x^2\!+\!(1\!-\!w^2)(1\!+\!\hat{\eta}^2h^2x^2)\right]}}\,dx~,~~~~~~~   \label{T-Define}
\end{eqnarray}
with
\begin{equation}
\frac{b}{r_0}=\frac{1}{w}\sqrt{\frac{\left(1+\hat{\eta}^2h^2\right)\left[w^2\sqrt{1+\hat{\eta}^2h^2}+2(1-w^2)h\right]}{\sqrt{1+\hat{\eta}^2h^2}-2h}} ~.  \label{b-r0-4}
\end{equation}
In order to obtain the analytical expression of $T$, we define $\zeta\equiv r_0/R$, make a power series expansion for the integrand of Eq.~\eqref{T-Define} in $h$, and finally integrate the expansion over $x$ to have
\begin{eqnarray}
T(R)=\frac{\sqrt{R^2-r_0^2}}{w}+\frac{hr_0}{w^3}\!\left[\!\frac{\sqrt{1-\zeta^2}}{1+\zeta}+(3w^2\!-\!1)\ln\!\left(\frac{1+\sqrt{1-\zeta^2}}{\zeta}\right)\!\right]
\!+\!\frac{h^2r_0V_{15}}{2w}+\frac{h^3r_0V_{16}}{2w}+\mathcal{O}\left(M^4\right)~, \label{T-Series}
\end{eqnarray}
where
\begin{eqnarray}
&&V_{15}=\left(15+\hat{\eta}^2\right)\left(\frac{\pi}{2}-\arcsin{\zeta}\right)
-\frac{\sqrt{1-\zeta^2}}{w^2\left(1+\zeta\right)}\!\left[6-\frac{2+\zeta}{w^2(1+\zeta)}\right]~,~                \label{V15}  \\
&&V_{16}=\frac{15}{w^2}\!\left(\arcsin{\zeta}\!-\!\frac{\pi}{2}\right)
\!+\!\left[35\!+\!\frac{23}{w^2(1+\zeta)}\!+\!\frac{3(1+2\zeta)}{w^4(1+\zeta)^2}\!-\!\frac{1+3\zeta+\zeta^2}{w^6(1+\zeta)^3}\right]\!\!\sqrt{1\!-\!\zeta^2}
+\frac{\hat{\eta}^2}{w^2}\!\left(\arcsin{\zeta}\!-\!\frac{\pi}{2}\!+\!\frac{2\sqrt{1\!-\!\zeta^2}}{1+\zeta}\,\right)~.~~~~~~    \label{V16}
\end{eqnarray}

It is convenient to express $T(R)$ presented in Eq.~\eqref{T-Series} in terms of the impact parameter $b$, for which the evaluation of the magnitudes of order for $M/b$ and $b/R$ is necessary. Let $R_S$ and $R_O$ denote the radial coordinates of the source $S$ and the observer $O$, respectively. Then, we have $R_S=\left(d_{LS}^2+d_{S}^2\tan^2\mathcal{B}\right)^{1/2}$ and $R_O=d_L$. By noticing $M/b\sim\varepsilon$,
$b/R_S\sim D(1\!-\!D)\left(D^2+\tan^2\mathcal{B}\right)^{-1/2}\varepsilon$, and $b/R_O\sim D\varepsilon$~\cite{KP2005}, we further expand Eq.~\eqref{T-Series} in $\varepsilon$ as
\begin{equation}
\frac{T(R)}{R}=\frac{1}{w}-\frac{1}{2w}\frac{b}{R}\!\left[\frac{b}{R}-\frac{2}{w^2}\frac{M}{b}+2\left(3-\frac{1}{w^2}\right)\frac{M}{b}\ln\!\left(\frac{b}{2R}\right)\right]
+\frac{\pi(15+\hat{\eta}^2)}{4w}\frac{b}{R}\frac{M^2}{b^2}+\mathcal{O}(\varepsilon^4)~.   \label{T-Series-2}
\end{equation}

Based on Eq.~\eqref{T-Series-2}, the gravitational time delay for a massive particle travelling from the source to the observer is expressed as
\begin{equation}
\tau=T(R_S)+T(R_O)-\frac{d_S}{w\cos\mathcal{B}}~,   \label{T-total}
\end{equation}
which can be reduced to its lightlike definition~\cite{KP2005} for the case of $w=1$. To compare with astronomical measurements, we apply $b=d_L\sin\vartheta$ and $M=d_L\tan\vartheta_{\bullet}$
to Eq.~\eqref{T-total} and get
\begin{eqnarray}
\tau=\frac{8d_Ld_{LS}}{w^3d_S}\bigg{\{}\!\left[1\!+\!w^2\beta^2\!-\!w^2\theta_0^2\!+\!\frac{1\!-\!3w^2}{2}\ln\left(\frac{d_L\theta_0^2\vartheta_E^2}{4d_{LS}}\right)\!\right]\!\varepsilon^2
\!+\!\frac{\pi w^2\left(15\!+\!\hat{\eta}^2\right)\!+\!4\left(1\!-\!3w^2\!-\!2w^2\theta_0^2\right)\theta_1}{4\theta_0}\,\varepsilon^3\!+\!\mathcal{O}\left(\varepsilon^4\right)\!\bigg{\}}~.~~~~~~~  \label{T-Series-3}
\end{eqnarray}
The substitution of Eq.~\eqref{theta1} into Eq.~\eqref{T-Series-3} and the use of the natural time scale $\tau_E\equiv d_L\vartheta_E^2/D \left(=4M\right)$ yield the following scaled gravitational time delay
\begin{eqnarray}
\nn&&\hat{\tau}\equiv\frac{\tau}{\tau_E}=\frac{1}{2w^3}\!\left[1+w^2\beta^2-w^2\theta_0^2
+\frac{1-3w^2}{2}\ln\left(\frac{d_L\theta_0^2\vartheta_E^2}{4d_{LS}}\right)\right]   \\
&&\hspace*{47pt}+\,\frac{\pi}{8w^3\theta_0}\!\left[w^2\left(15+\hat{\eta}^2\right)+\frac{3\left(4+w^2\right)
+w^2\hat{\eta}^2}{2}\frac{1-3w^2-2w^2\theta_0^2}{1+w^2+2w^2\theta_0^2}\right]\varepsilon+\mathcal{O}(\varepsilon^2)~.~~~~~~   \label{T-Series-4}
\end{eqnarray}
The difference between the scaled gravitational time delays of the primary and secondary images can be finally obtained from Eq.~\eqref{T-Series-4} as
\begin{equation}
\Delta\hat{\tau}=\hat{\tau}_{-}-\hat{\tau}_{+}=\Delta\hat{\tau}_0+\Delta\hat{\tau}_1\varepsilon+\mathcal{O}(\varepsilon^2)~, \label{Differential-TD}
\end{equation}
where
\begin{eqnarray}
&&\Delta\hat{\tau}_0=\frac{(\theta_0^{+})^2-(\theta_0^{-})^2}{2w}+\frac{1-3w^2}{2w^3}\ln\left(\frac{\theta_0^{-}}{\theta_0^{+}}\right)~,~~  \label{Delta-1} \\
&&\Delta\hat{\tau}_1=\frac{\pi}{8w^3}\!\left\{w^2(15+\hat{\eta}^2)\!\left(\frac{1}{\theta_0^{-}}\!-\!\frac{1}{\theta_0^{+}}\right)
\!+\!\frac{3(4\!+\!w^2)\!+\!w^2\hat{\eta}^2}{2}\!\left\{\frac{1-3w^2-2w^2(\theta_0^{-})^2}{\theta_0^{-}[1+w^2+2w^2(\theta_0^{-})^2]}
-\frac{1-3w^2-2w^2(\theta_0^{+})^2}{\theta_0^{+}[1+w^2+2w^2(\theta_0^{+})^2]}\right\} \right\}~,~~~~~~~  \label{Delta-2}
\end{eqnarray}
or equivalently in terms of the source position,
\begin{eqnarray}
&&\Delta\hat{\tau}_0=\frac{\xi\left|\beta\right|}{2w}
+\frac{1-3w^2}{2w^3}\ln\!\left(\frac{\xi-\left|\beta\right|}{\xi+\left|\beta\right|}\right)~, \label{Delta-1-beta} \\
&&\Delta\hat{\tau}_1=\frac{\pi\left[12-3w^2(1-\hat{\eta}^2)+w^4(21-\hat{\eta}^2)\right] \left|\beta\right|}{8w(1+w^2)^2} ~. ~~~~~~ \label{Delta-2-beta}
\end{eqnarray}
The values of the coefficients of the zeroth- and first-order contributions to $\Delta\hat{\tau}$ are shown in Fig.\,\ref{Figure3}. It should be stressed that Eqs.~\eqref{T-Series} and \eqref{Differential-TD} are consistent with the lightlike results~\cite{CX2021,ZX2022} when $w=1$ is assumed. Additionally, those equations agree with the results of the Kerr-Newman lensing of massive particles given in~\cite{HL2022} for the case of no bounce parameter, and can be simplified to the null results in Schwarzschild spacetime~\cite{KP2005} in the limit $w\rightarrow1$ and $\eta\rightarrow0$. Moreover, according to Eq.~\eqref{Delta-2-beta}, it is found that the leading-order contribution from the bounce parameter to the scaled differential time delay of timelike signals is always positive, which may also be used to measure or constrain the bounce parameter itself.

\section{Discussion of detectable quantities} \label{sect5}

\subsection{A brief review}
The conventional weak-field lensing observables include the positions, fluxes, and the differential time delay of the images of a light source, if they are resolvable. Otherwise the total flux and the centroid position are the main practical observables. It is worth mentioning that the proper combinations of these lensing quantities are interesting and may also serve as important observables~\cite{KP2006a,WP2007}, and it should be better to contain the sum and difference relations of the fluxes and positions of the primary and secondary images of the light source (see Section~\ref{sect4}) when considering practical observable quantities. Notice that the image flux $F\left(=\Sigma F_i\varepsilon^i\right)$ is related to the magnification by $F_i=|\mu_i|F_{\text{s}}$, where $i\in N$ and $F_{\text{s}}$ is the positive flux of unlensed light signals or unlensed massive particle signals~\cite{PNHZ2014,BT2017} emitted by the source.

As mentioned above, the deviation of the initial velocity $w$ of a timelike particle from the speed of light may have an influence on all kind of measurable gravitational effects. It is of great significance to discuss the velocity-induced effect caused by this deviation~\cite{WS2004,LYJ2016,HL2022}, which is the key difference between the gravitational lens effects of timelike and lightlike signals. Hence, our concentration for potential applications includes the velocity effects on these lensing observables of the images of a point source of light in the black-bounce-Schwarzschild spacetime, and the bounce-induced effects on the Schwarzschild lensing observables of the images of a point-like massive-particle source, given the fact that the weak- and strong-field gravitational lensing effects of light signals in the black-bounce-Schwarzschild spacetime have been investigated in detail. Note that these correctional effects are also detectable, and that the conversion from the convenient scaled quantities $\{\theta,~\beta,~\mu,~\Theta_{\text{cent}},~\hat{\tau}\}$ to the practical measurable quantities $\{\vartheta,~\mathcal{B},~F,~\Xi_{\text{cent}},~\tau\}$ becomes necessary for connecting with astronomical observations, with $\vartheta^{\pm}=\vartheta_E\hspace*{0.8pt}\theta^{\pm}$, $F^{\pm}=\pm F_{\text{s}}\hspace*{0.8pt}\mu^{\pm}$, and $\Xi_{\text{cent}}=\vartheta_E\hspace*{0.8pt}\Theta_{\text{cent}}$.

\subsection{Velocity-induced effects} \label{VA}
In terms of the physical quantities $\{\vartheta,~\mathcal{B},~F,~\Xi_{\text{cent}},~\tau\}$, the nonvanishing components of the velocity effects on the zeroth-, first-, or second-order practical lensing observables of the images of the light source in the black-bounce-Schwarzschild black hole spacetime can be expressed according to the results given in Section~\ref{sect4} as follows~\cite{HL2022}:
\begin{eqnarray}
&&\delta\vartheta_i^{\pm}\varepsilon^i\equiv\vartheta_E\left(\theta_i^{\pm}-\theta_i^{\pm}\!\left.\right|_{w=1}\right)\varepsilon^i~,~~~~~    \label{NS-1}  \\
&&\delta F_i^{\pm}\varepsilon^i\equiv\pm\,F_{\text{s}}\left(\mu_i^{\pm}-\mu_i^{\pm}\!\left.\right|_{w=1}\right)\varepsilon^i~,~~~~~        \label{NS-2}  \\
&&\delta(\vartheta_i^{+}+\vartheta_i^{-})\,\varepsilon^i\equiv\vartheta_E\left[(\theta_i^{+}+\theta_i^{-})-(\theta_i^{+}+\theta_i^{-})\!\left.\right|_{w=1}\right]\varepsilon^i~,  \label{NS-3} \\
&&\delta(\vartheta_j^{+}-\vartheta_j^{-})\,\varepsilon^j\equiv\vartheta_E\left[(\theta_j^{+}-\theta_j^{-})-(\theta_j^{+}-\theta_j^{-})\!\left.\right|_{w=1}\right]\varepsilon^j~,  \label{NS-4} \\
&&\delta(F_k^{+}+F_k^{-})\,\varepsilon^k\equiv F_{\text{s}}\left[(\mu_k^{+}-\mu_k^{-})-(\mu_k^{+}-\mu_k^{-})\!\left.\right|_{w=1}\right]\varepsilon^k~, \label{NS-5}  \\
&&\delta(F_l^{+}-F_l^{-})\,\varepsilon^l\equiv F_{\text{s}}\left[(\mu_l^{+}+\mu_l^{-})-(\mu_l^{+}+\mu_l^{-})\!\left.\right|_{w=1}\right]\varepsilon^l ~, ~~~~  \label{NS-6}  \\
&&\delta\Xi_{\text{cent,}k}\,\varepsilon^k\equiv\vartheta_E\left(\Theta_{\text{cent,}k}-\Theta_{\text{cent,}k}\!\left.\right|_{w=1}\right)\varepsilon^k~,     \label{NS-7}  \\
&&\delta\Delta\tau_m\,\varepsilon^m\equiv\tau_E\left(\Delta\hat{\tau}_m-\Delta\hat{\tau}_m\!\left.\right|_{w=1} \right)\varepsilon^m~,   \label{NS-8}
\end{eqnarray}
with $i\in\{0,~1,~2\}$, $j\in\{1,~2\}$, $k\in\{0,~2\}$, $l\in\{1\}$, and $m\in\{0,~1\}$ here and in the following of this section. Concretely, $\delta\vartheta_i^{\pm}\varepsilon^i$ $($or $\delta F_i^{\pm}\varepsilon^i)$ denote the velocity effects on the $i\,$th-order unscaled angular positions (or on the $i\,$th-order fluxes) of the primary and secondary images of the light source, respectively. $\delta(\vartheta_i^{+}+\vartheta_i^{-})\,\varepsilon^i$ and $\delta(\vartheta_j^{+}-\vartheta_j^{-})\,\varepsilon^j$ represent the velocity effects on the $i\,$th-order practical positional sum and on the $j\,$th-order unscaled positional difference of the light-source images, respectively. $\delta(F_k^{+}+F_k^{-})\,\varepsilon^k$ and $\delta(F_l^{+}-F_l^{-})\,\varepsilon^l$ denote respectively the velocity effects on the $k\,$th-order flux sum and on the $l\,$th-order flux difference of the images of the light source. Finally, the velocity effects on the $k\,$th-order unscaled magnification centroid and on the $m\,$th-order unscaled differential time delay of the light-source images are denoted by $\delta\Xi_{\text{cent,}k}\,\varepsilon^k$ and $\delta\Delta\tau_m\,\varepsilon^m$, respectively. Additionally, the following relations hold well in our lensing scenario:
\begin{eqnarray}
&&\delta(\vartheta_0^{+}+\vartheta_0^{-})=2\delta\vartheta_0^{+}~,                                              \label{SR1} \\
&&\delta(F_0^{+}+F_0^{-})=2\delta F_0^{+}~,                                           \label{SR2} \\
&&\delta(F_1^{+}-F_1^{-})\,\varepsilon=2\delta F_1^{+}\varepsilon~,                   \label{SR3} \\
&&\delta(F_2^{+}+F_2^{-})\,\varepsilon^2=2\delta F_2^{+}\varepsilon^2~,               \label{SR4} \\
&&\delta(F_0^{+}-F_0^{-})=\delta(F_1^{+}+F_1^{-})\,\varepsilon=\delta(F_2^{+}-F_2^{-})\,\varepsilon^2=0~.~~~~ \label{SR5}
\end{eqnarray}

Furthermore, it is well-known that magnitude is used to indicate the brightness of a source conventionally and is related to the received flux of electromagnetic waves via the Pogson formula~\cite{IMKTP2018}. However, for a more convenient discussion of the deviations from the lightlike counterparts, we may convert the velocity effects on the individual received fluxes and on the sum and difference relations of the fluxes of the light-source images into the following magnitude-like differences:
\begin{eqnarray}
&&\delta m_1^{\pm}\equiv-2.5\lg\!\left[1+\frac{\delta F^{\pm}}{F^{\pm}\!\!\left.\right|_{w=1}}\right]
=-2.5\lg\!\left[1+\frac{\sum_{i=0}^2 \delta F_i^{\pm}\varepsilon^i}{\sum_{i=0}^2 F_i^{\pm}\!\!\left.\right|_{w=1}\varepsilon^i}+\mathcal{O}(\varepsilon^3)\right] ~,   \label{M-1} \\
&&\delta m_2\equiv-2.5\lg\!\left[1+\frac{\delta\left(F^{+}+F^{-}\right)}{\left(F^{+}+F^{-}\right)\!\left.\right|_{w=1}}\right]
=-2.5\lg\!\left[1\!+\!\frac{\delta(F_0^{+}+F_0^{-})+\delta(F_2^{+}+F_2^{-})\,\varepsilon^2}
{(F_0^{+}+F_0^{-})\!\!\left.\right|_{w=1}+(F_2^{+}+F_2^{-})\!\!\left.\right|_{w=1}\varepsilon^2}+\mathcal{O}(\varepsilon^3)\right]~,   \label{M-2}  \\
\hspace*{-1cm}&&\delta m_3\equiv-2.5\lg\!\left[1+\frac{\delta\left(F^{+}-F^{-}\right)}{\left(F^{+}-F^{-}\right)\!\left.\right|_{w=1}}\right]
=-2.5\lg\!\left[1+\frac{\delta(F_1^{+}-F_1^{-})\,\varepsilon}{F_{\text{s}}+(F_1^{+}-F_1^{-})\!\!\left.\right|_{w=1}\varepsilon}+\mathcal{O}(\varepsilon^3)\right]~,  \label{M-3}
\end{eqnarray}
where $\delta m_1^{\pm}$ denote the magnitude differences induced by the velocity effects on the fluxes of the primary and secondary images of the light source, respectively, while $\delta m_2$ and $\delta m_3$ are used to represent respectively the differential magnitudes caused by the velocity effects on the sum and difference relations of the fluxes of the light-source images. In the following section, the magnitude differences originated from the velocity effects on the $i$\,th-order contributions to $F^{+}$, $F^{+}+F^{-}$, and $F^{+}-F^{-}$ are further denoted by $(\delta m_1^{+})_{i\text{\,th}}$, $(\delta m_2)_{i\text{\,th}}$, and $(\delta m_3)_{i\text{\,th}}$, respectively.

\subsection{Black-bounce-induced effects}
Since the influences of the bounce parameter of the spacetime on the lensing observables of the images of the point-like massive-particle source are a part of our main focus, the bounce-induced effects on the measurable Schwarzschild lensing properties (i.e., the deviations of the lensing properties of massive particles in black-bounce-Schwarzschild black hole geometry from those in Schwarzschild spacetime) also deserve to be considered. Similarly, the nonzero bounce-induced effects on the $j\,$th-order unscaled positions or on the $j\,$th-order fluxes of the primary and secondary images of the particle source, together with these on the sum and difference relations of the $j\,$th-order unscaled positions of the particle-source images, on the $n\,$th-order flux sum and $l\,$th-order flux difference of the images of the particle source, on the $n\,$th-order unscaled magnification centroid, and on the $l\,$th-order differential time delay of the particle-source images, are given in terms of the quantities $(\vartheta,~\mathcal{B},~F,~\Xi_{\text{cent}},~\tau)$ as follows:
\begin{eqnarray}
&&\Delta\vartheta_j^{\pm}\varepsilon^j\equiv\vartheta_E\left(\theta_j^{\pm}-\theta_j^{\pm}\!\left.\right|_{\hat{\eta}=0}\right)\varepsilon^j~,    \label{NS-1-BB}  \\
&&\Delta F_j^{\pm}\varepsilon^j\equiv\pm\,F_{\text{s}}\left(\mu_j^{\pm}-\mu_j^{\pm}\!\left.\right|_{\hat{\eta}=0}\right)\varepsilon^j~,         \label{NS-2-BB}  \\
&&\Delta(\vartheta_j^{+}\pm\vartheta_j^{-})\,\varepsilon^j\equiv\vartheta_E\left[(\theta_j^{+}\pm\theta_j^{-})-(\theta_j^{+}\pm\theta_j^{-})\!\left.\right|_{\hat{\eta}=0}\right]\varepsilon^j~,  \label{NS-3-BB} \\
&&\Delta(F_n^{+}+F_n^{-})\,\varepsilon^n\equiv\,F_{\text{s}}\left[(\mu_n^{+}-\mu_n^{-})-(\mu_n^{+}-\mu_n^{-})\!\left.\right|_{\hat{\eta}=0}\right]\varepsilon^n~, ~~~~  \label{NS-5-BB}  \\
&&\Delta(F_l^{+}-F_l^{-})\,\varepsilon^l\equiv\,F_{\text{s}}\left[(\mu_l^{+}+\mu_l^{-})-(\mu_l^{+}+\mu_l^{-})\!\left.\right|_{\hat{\eta}=0}\right]\varepsilon^l~, ~~~~  \label{NS-6-BB}  \\
&&\Delta\Xi_{\text{cent,}n}\,\varepsilon^n\equiv\vartheta_E\left(\Theta_{\text{cent,}n}-\Theta_{\text{cent,}n}\!\left.\right|_{\hat{\eta}=0}\right)\varepsilon^n~,     \label{NS-7-BB}  \\
&&\Delta(\Delta\tau_{\hspace{1pt}l})\,\varepsilon^l\equiv\tau_E\left(\Delta\hat{\tau}_{\hspace{1pt}l}-\Delta\hat{\tau}_{\hspace{1pt}l}\!\left.\right|_{\hat{\eta}=0} \right)\varepsilon^l~,   \label{NS-8-BB}
\end{eqnarray}
with $n\in\{2\}$. Note that $\Delta(\vartheta_1^{+}+\vartheta_1^{-})\,\varepsilon$ is independent on $\beta$ and that we have $\Delta(F_1^{+}-F_1^{-})\varepsilon=2\Delta F_1^+\varepsilon$ and $\Delta(F_2^{+}+F_2^{-})\varepsilon^2=2\Delta F_2^{+}\varepsilon^2$, in view of $\mu_1^+=\mu_1^-$ and $\mu_2^+=-\mu_2^-$. In a similar way, the bounce effects on the fluxes and on the flux sum and difference relations of the images of the particle source may also be converted into several magnitude-like differences, which are defined by
\begin{eqnarray}
&&\Delta m_1^{\pm}\equiv-2.5\lg\!\left[1+\frac{\Delta F^{\pm}}{F^{\pm}\!\!\left.\right|_{\hat{\eta}=0}}\right]
=-2.5\lg\!\left[1+\frac{\sum_{j=1}^2\Delta F_j^{\pm}\varepsilon^j}{\sum_{i=0}^2 F_i^{\pm}\!\!\left.\right|_{\hat{\eta}=0}\varepsilon^i }+\mathcal{O}(\varepsilon^3)\right] ~,   \label{M-4} \\
&&\Delta m_2\equiv-2.5\lg\!\left[1\!+\!\frac{\Delta\left(F^{+}+F^{-}\right)}{\left(F^{+}\!+\!F^{-}\right)\!\left.\right|_{\hat{\eta}=0}}\right]
=-2.5\lg\!\left[1\!+\!\frac{\Delta(F_2^{+}+F_2^{-})\,\varepsilon^2}
{(F_0^{+}\!+\!F_0^{-})\!\!\left.\right|_{\hat{\eta}=0}+(F_2^{+}\!+\!F_2^{-})\!\!\left.\right|_{\hat{\eta}=0}\varepsilon^2}+\mathcal{O}(\varepsilon^3)\right]~,~~~~~~~  \label{M-5}  \\
\hspace*{-1cm}&&\Delta m_3\equiv-2.5\lg\!\left[1+\frac{\Delta \left(F^{+}-F^{-}\right)}{\left(F^{+}-F^{-}\right)\!\left.\right|_{\hat{\eta}=0}}\right]
=-2.5\lg\!\left[1+\frac{\Delta(F_1^{+}-F_1^{-})\,\varepsilon}{F_{\text{s}}+(F_1^{+}-F_1^{-})\!\!\left.\right|_{\hat{\eta}=0}\varepsilon}+\mathcal{O}(\varepsilon^3)\right]~.~~~~~~~~~  \label{M-6}
\end{eqnarray}
\end{widetext}
Here $\Delta m_1^{\pm}$ are, respectively, the differential magnitudes originated from the bounce effects on the fluxes of the primary and secondary images of the particle source. $\Delta m_2$ and $\Delta m_3$ stand for the differential magnitudes due to the bounce effects on the flux sum and difference relations of the particle-source images, respectively. Also, we use $(\Delta m_1^{+})_{j\text{\,th}}$, $(\Delta m_2)_{j\text{\,th}}$, and $(\Delta m_3)_{j\text{\,th}}$ to denote respectively the magnitude differences due to the bounce effects on the $j$\,th-order contributions to $F^{+}$, $F^{+}+F^{-}$, and $F^{+}-F^{-}$.

\begin{figure*}
\centering
\begin{minipage}[b]{13cm}
\includegraphics[width=13cm]{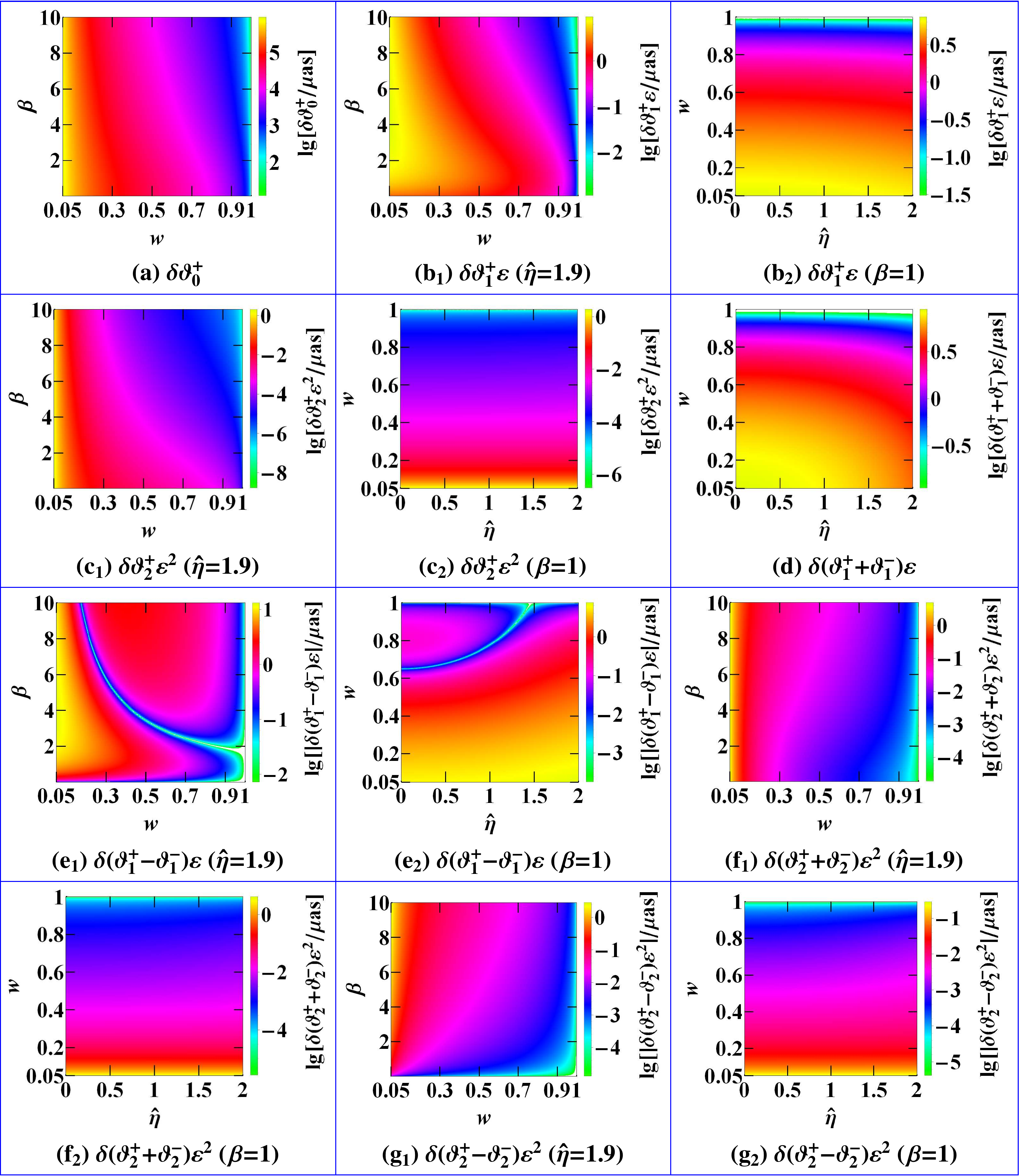}
\end{minipage}
\caption{The color-indexed velocity effects on the zeroth-, first-, and second-order contributions to the unscaled angular position $\vartheta^{+}$ of the primary image of the light source, as well as those on the first- and second-order unscaled positional sum and difference of the light-source images, plotted as the functions of $w$ and $\beta$ or $\hat{\eta}$ and $w$ (on the domain $\mathcal{D}_{\hat{\eta}1}$ or $\mathcal{D}_{\beta}$). Note that $\delta(\vartheta_0^{+}+\vartheta_0^{-})$ is not plotted due to the relation $\delta(\vartheta_0^{+}+\vartheta_0^{-})=2\delta\vartheta_0^{+}$. For comparison, the figure of $\delta\vartheta_0^{+}$ is taken from~\cite{HL2022}.  } \label{Figure4}
\end{figure*}

\section{Application to the galactic supermassive black hole} \label{sect6}
An interesting application of our analytical results is the consideration of the scenario where the galactic supermassive black hole, Sgr A$^{\ast}$, is modeled as a black-bounce-Schwarzschild lens. In view of the detailed investigation on the lensing scenarios where Sgr A$^{\ast}$ is regarded as a Schwarzschild~\cite{VE2000,KP2005} or black-bounce-Schwarzschild~\cite{NPPS2020,CX2021,ZX2022} lens for photons being the test particles, in this section the discussion focuses concretely on (i) the velocity-induced effects on the practical (or unscaled) lensing observables of the images of the light source, (ii) the bounce-induced effects on the measurable Schwarzschild lensing properties of the images of the massive-particle source, and (iii) the possibilities of their astronomical measurements. Concerning the last aspect, it should be mentioned that some challenges~\cite{MFHM2019,Markus2019,Frost2023} owing to the extension from the astronomical observations via multiwavelength photons to the observations via non-photonic messengers yield a foreseeable result that current astronomical observatories or instruments observing massive particles or multi-messengers are relatively rare, with lower (or even much lower) angular, flux, and time resolutions (see, e.g.,~\cite{IceCube2006,Aab2014,Budnev2016,BT2017,DEGG2021}). Thus, we may base on the capabilities of current multi-frequency astronomical surveys~\cite{RD2020} and particle detectors or observatories, and perform a rough discussion regarding the potential observations of the velocity-induced and bounce-induced effects, both of which are the deviations of the routine lensing properties in different scenarios.

We know that Sgr A$^{\ast}$ has a mass $M=4.2\times10^6M_{\odot}$~\cite{BG2016,Parsa2017} and a distance $d_L=8.2$\,kpc~\cite{BG2016} from us, where $M_{\odot}$ ($=1.475$\,km) denotes the mass of the Sun.
It means that the angular gravitational radius $\vartheta_{\bullet}$ and the lensing time scale $\tau_E$ are $5.06\,\mu$as and $82.6\,$s, respectively. Additionally, the source of massive particles is assumed to be at a distance $d_{LS}=0.01$\,kpc~\cite{HL2022} ($d_{LS}\ll d_L\approx d_S$) from the central black hole. Thus, the parameter $D$, the angular Einstein radius $\vartheta_E$, and the expansion parameter $\varepsilon$ have values $1.22\times10^{-3}$, $0.071$\,as, and $7.12\times10^{-5}$, respectively. As illustrated in Section~\ref{IP}, $0.05\lesssim w\leq1$ and $0<\hat{\eta}<2$ are also supposed in our discussion, since we only consider massive (and massless) test particles which are relativistic. Moreover, we take the primary image as an example to analyze the image properties and assume $0.01\leq\beta\leq10$~\cite{CX2021}. The following exemplary domains are adopted to perform the evaluation:
\begin{widetext}
\begin{eqnarray}
&&\mathcal{D}_{\hat{\eta}1}\equiv\left\{(w,~\beta,~\hat{\eta})\,|\,0.05\lesssim w\leq1,~0.01\leq\beta\leq10,~\hat{\eta}=1.9\right\}~,   \label{D-1} \\
&&\mathcal{D}_{\hat{\eta}2}\equiv\left\{(w,~\beta,~\hat{\eta})\,|\,0.05\lesssim w\leq1,~0.01\leq\beta\leq10,~\hat{\eta}=0.1\right\}~,   \label{D-2} \\
&&\mathcal{D}_{\beta}\equiv\left\{(w,~\beta,~\hat{\eta})\,|\,0.05\lesssim w\leq1,~\beta=1,~0<\hat{\eta}<2\right\}~,   \label{D-3} \\
&&\mathcal{D}_{w1}\equiv\left\{(w,~\beta,~\hat{\eta})\,|\,w=0.5,~0.01\leq\beta\leq10,~0<\hat{\eta}<2\right\}~,   \label{D-4} \\
&&\mathcal{D}_{w2}\equiv\left\{(w,~\beta,~\hat{\eta})\,|\,w=1,~0.01\leq\beta\leq10,~0<\hat{\eta}<2\right\}~.   \label{D-5}
\end{eqnarray}
\end{widetext}

\subsection{Velocity effects on the practical lensing observables}
We first discuss the velocity effects on the angular positions and on the positional relations of the images of the light source. Figure~\ref{Figure4} shows $\delta\vartheta_0^+,~\delta\vartheta_1^+\varepsilon,~\delta\vartheta_2^+\varepsilon^2,~\delta(\vartheta_1^{+}\pm\vartheta_1^{-})\,\varepsilon$, and $\delta(\vartheta_2^{+}\pm\vartheta_2^{-})\,\varepsilon^2$ for various $\beta$ (or $\hat{\eta}$) and $w$ in color-indexed form, and their values (except the ones for $\delta\vartheta_0^+$~\cite{HL2022}) are given in Tab.~\ref{Table1}. On the one hand, the dependence of these velocity effects on the variables deserves to be analyzed qualitatively. Similar to the behavior of the velocity effect on the zeroth-order unscaled position of the light-source primary image~\cite{HL2022}, the velocity effects on the first- and second-order contributions to $\vartheta^+$, together with those on the first- and second-order unscaled positional sum relations, decrease monotonically with the increase of the particle velocity $w$ on its domain for any given source position $\beta\in\left[0.01,~10\right]$ and a fixed bounce parameter $\hat{\eta}=1.9$. The same tendency of change of these four velocity effects is obtained for any given bounce parameter $\hat{\eta}\in\left(0,~2\right)$, when $\beta=1$ is assumed. This is not the case for the velocity effects on the first- and second-order unscaled positional difference relations. Figure~\ref{Figure4} and Table~\ref{Table1} indicate that $\delta(\vartheta_2^{+}-\vartheta_2^{-})\,\varepsilon^2$ has a reverse tendency of change on the domains of $\mathcal{D}_{\hat{\eta}1}$ and $\mathcal{D}_{\beta}$. Differently, $\delta(\vartheta_1^{+}-\vartheta_1^{-})\,\varepsilon$ decreases to a minimum value firstly and then increases to zero when increasing $w$ for the domains $\left\{(w,~\beta,~\hat{\eta})\,|\,0.05\lesssim w\leq1,~1.72\lesssim\beta\leq10,~\hat{\eta}=1.9\right\}$ and $\left\{(w,~\beta,~\hat{\eta})\,|\,0.05\lesssim w\leq1,~\beta=1,~0<\hat{\eta}\lesssim1.46\right\}$, while it decreases from a maximum value to zero monotonically with the increase of $w$ for the domains $\left\{(w,~\beta,~\hat{\eta})\,|\,0.05\lesssim w\leq1,~0.01\leq\beta\lesssim1.72,~\hat{\eta}=1.9\right\}$ and $\left\{(w,~\beta,~\hat{\eta})\,|\,0.05\lesssim w\leq1,~\beta=1,~1.46\lesssim\hat{\eta}<2\right\}$. It should be pointed out that the minimum and maximum values of $\delta(\vartheta_1^{+}-\vartheta_1^{-})\,\varepsilon$ may vary with the change of the given values of $\beta$ and $\hat{\eta}$. On the other hand, we perform a brief analysis on the possibilities of detecting these velocity effects on the basis of the results given in Fig.~\ref{Figure4} and Tab.~\ref{Table1}. Although smaller than that of detecting $\delta\vartheta_0^{+}$ whose value is much larger than current multi-wavelength astrometric precision (at the tens of $\mu$as level or better~\cite{Brown2021,Abuter2017a}) for any given source position $\beta\in\left[0.01,~10\right]$ and most of relativistic massive particles with $0.05\lesssim w\lesssim0.99$ (see Tab.\,I of~\cite{HL2022} for details), the possibility to detect the velocity effect on the first-order sum relation of the practical positions of the light-source images in (near) future is relatively large. For example, even for a massive particle with a high relativistic velocity $w=0.98$ and a large bounce parameter $\hat{\eta}=1.99$, the value of $\delta(\vartheta_1^{+}+\vartheta_1^{-})\varepsilon$ can reach about $0.1\mu$as, which is larger than the intended accuracy ($\sim\!0.05\mu$as) of the proposed NEAT mission~\cite{Malbet2012,Malbet2014}. Under this case, $\delta(\vartheta_1^{+}+\vartheta_1^{-})\varepsilon$ may still be measured by future joint multi-messenger detectors whose angular resolution is approximately equal to or better than that of NEAT. For any combination of $w\in\left[0.05,~0.99\right]$ and $\hat{\eta}\in\left(0,~1.999\right]$, $\delta(\vartheta_1^{+}+\vartheta_1^{-})\varepsilon$ ranges from about $0.05\mu$as to a maximum value $8.89\mu$as which is independent on the angular source position. Moreover, it is likely to measure the velocity effect on the first-order practical position of the primary image or on the first-order unscaled positional difference of the images of the light source in most cases of the domains $\mathcal{D}_{\hat{\eta}1}$ and $\mathcal{D}_{\beta}$ within the capabilities of these NEAT-level multi-messenger detectors, though the upper limit of $w$ has to be restricted to a lower value. For instance, this upper limit is about 0.83 and 0.98 for $\delta\vartheta_1^{+}\varepsilon$ on $\mathcal{D}_{\hat{\eta}1}$ and $\mathcal{D}_{\beta}$, respectively. There is a relatively small possibility of detecting the velocity effect on the second-order contribution to $\vartheta^{+}$ or to the sum relation of the practical positions via these future multi-messenger detectors. It is not until the test particle travels with a limited relativistic velocity that the values of $\delta\vartheta_2^{+}\varepsilon^2$ and $\delta(\vartheta_2^{+}+\vartheta_2^{-})\,\varepsilon^2$ can be larger than $0.05\mu$as. On the domain $\mathcal{D}_{\beta}$, the approximate ranges $0.05\lesssim w\lesssim0.16$ and $0.05\lesssim w\lesssim0.23$ should be satisfied for $\delta\vartheta_2^{+}\varepsilon^2$ and $\delta(\vartheta_2^{+}+\vartheta_2^{-})\,\varepsilon^2$, respectively, to get a possible detection by the NEAT-level multi-messenger detectors. The required ranges become $0.05\lesssim w\lesssim0.1$ and $0.05\lesssim w\lesssim0.23$ for $\delta\vartheta_2^{+}\varepsilon^2$ and $\delta(\vartheta_2^{+}+\vartheta_2^{-})\,\varepsilon^2$ on $\mathcal{D}_{\hat{\eta}1}$, respectively. Concerning the velocity effect on the second-order unscaled positional difference relation, it may be measured by these multi-messenger detectors only when $w$ takes a small relativistic value and the combination of $\beta$ and $\hat{\eta}$ is proper. However, it is interesting to find there exist a few cases where the value of $\delta(\vartheta_2^{+}-\vartheta_2^{-})\,\varepsilon^2$ can be larger than $1\mu$as and may thus be detected by means of the mentioned multi-messenger detectors. For instance, its absolute value is about $2.9\mu$as, if $w=0.05,~\hat{\eta}=1.9$, and $\beta=10$.

\begin{widetext}

\begin{table}
\begin{minipage}[t]{0.48\textwidth}
  \centering
\begin{tabular}{cccccccc} \toprule[1.2px]
      $\beta\:\backslash\:w$    &    ~~~~$ 0.05 $ ~~~~   &     ~~~~$0.1$~~~~       &     ~~~~$0.5$~~~~        &     ~~~~$0.9$~~~~    &           $0.99$               \\   \midrule[0.5pt] \vspace*{-8pt}  \\
                   0.01         &          2.70          &          2.66           &          1.62            &          0.28        &       $\vartriangle$           \\
                    0.1         &          3.08          &          3.00           &          1.72            &          0.29        &       $\vartriangle$           \\
                    0.5         &          4.69          &          4.45           &          2.15            &          0.33        &       $\vartriangle$           \\
                      1         &          6.37          &          5.92           &          2.47            &          0.34        &       $\vartriangle$           \\
                      5         &          9.17          &          7.26           &          1.02            &          0.09        &       $\vartriangle$           \\
                     10         &          7.76          &          4.86           &          0.33            &     $\vartriangle$   &       $\vartriangle$           \\   \bottomrule[1.2px]
\end{tabular} \par  \vspace*{3pt}
     \centerline{(a$_1$) $\delta\vartheta_1^{+}\varepsilon$ $\left(\hat{\eta}=1.9\right)$} \vspace*{10pt}
\end{minipage}
\begin{minipage}[t]{0.48\textwidth}
  \centering
\begin{tabular}{cccccccc} \toprule[1.2px]
 $\hat{\eta}\:\backslash\:w$    &    ~~~~$ 0.05 $ ~~~~   &     ~~~~$0.1$~~~~       &     ~~~~$0.5$~~~~        &    ~~~~$0.9$~~~~    &            $0.99$              \\   \midrule[0.5pt] \vspace*{-8pt}  \\
                   0.01         &          7.35          &          6.88           &          2.96            &         0.41        &       $\vartriangle$           \\
                    0.1         &          7.35          &          6.88           &          2.96            &         0.41        &       $\vartriangle$           \\
                    0.5         &          7.29          &          6.81           &          2.92            &         0.41        &       $\vartriangle$           \\
                      1         &          7.08          &          6.61           &          2.82            &         0.39        &       $\vartriangle$           \\
                    1.9         &          6.37          &          5.92           &          2.47            &         0.34        &       $\vartriangle$           \\
                  1.999         &          6.27          &          5.82           &          2.41            &         0.33        &       $\vartriangle$           \\   \bottomrule[1.2px]
\end{tabular} \par  \vspace*{3pt}
     \centerline{(a$_2$) $\delta\vartheta_1^{+}\varepsilon$ $\left(\beta=1\right)$}  \vspace*{10pt}
\end{minipage}

\begin{minipage}[t]{0.48\textwidth}
  \centering
\begin{tabular}{cccccccc} \toprule[1.2px]
      $\beta\:\backslash\:w$    &    ~~~~$ 0.05 $ ~~~~   &     ~~~~$0.1$~~~~       &      ~~~~$0.5$~~~~       &    ~~~~$0.9$~~~~    &            $0.99$              \\   \midrule[0.5pt] \vspace*{-8pt}  \\
                   0.01         &          2.06          &          0.27           &     $\vartriangle$       &   $\vartriangle$    &       $\vartriangle$           \\
                    0.1         &          2.05          &          0.27           &     $\vartriangle$       &   $\vartriangle$    &       $\vartriangle$           \\
                    0.5         &          1.99          &          0.25           &     $\vartriangle$       &   $\vartriangle$    &       $\vartriangle$           \\
                      1         &          1.92          &          0.24           &     $\vartriangle$       &   $\vartriangle$    &       $\vartriangle$           \\
                      5         &          1.43          &          0.13           &     $\vartriangle$       &   $\vartriangle$    &       $\vartriangle$           \\
                     10         &          0.97          &          0.06           &     $\vartriangle$       &   $\vartriangle$    &       $\vartriangle$           \\   \bottomrule[1.2px]
\end{tabular} \par  \vspace*{3pt}
     \centerline{(b$_1$) $\delta\vartheta_2^{+}\varepsilon^2$ $\left(\hat{\eta}=1.9\right)$ } \vspace*{10pt}
\end{minipage}
\begin{minipage}[t]{0.49\textwidth}
  \centering
\begin{tabular}{cccccccc} \toprule[1.2px]
 $\hat{\eta}\:\backslash\:w$    &    ~~~~$ 0.05 $ ~~~~   &     ~~~~$0.1$~~~~       &     ~~~~$0.5$~~~~        &     ~~~~$0.9$~~~~   &        $0.99$               \\   \midrule[0.5pt] \vspace*{-8pt}  \\
                   0.01         &          1.92          &          0.23           &    $\vartriangle$        &    $\vartriangle$   &    $\vartriangle$           \\
                    0.1         &          1.92          &          0.23           &    $\vartriangle$        &    $\vartriangle$   &    $\vartriangle$           \\
                    0.5         &          1.92          &          0.23           &    $\vartriangle$        &    $\vartriangle$   &    $\vartriangle$           \\
                      1         &          1.92          &          0.23           &    $\vartriangle$        &    $\vartriangle$   &    $\vartriangle$           \\
                    1.9         &          1.92          &          0.24           &    $\vartriangle$        &    $\vartriangle$   &    $\vartriangle$           \\
                  1.999         &          1.92          &          0.24           &    $\vartriangle$        &    $\vartriangle$   &    $\vartriangle$           \\   \bottomrule[1.2px]
\end{tabular} \par  \vspace*{3pt}
     \centerline{(b$_2$) $\delta\vartheta_2^{+}\varepsilon^2$ $\left(\beta=1\right)$ }
\end{minipage}

\begin{minipage}[t]{0.49\textwidth}
  \centering
\begin{tabular}{cccccccc} \toprule[1.2px]
 $\hat{\eta}\:\backslash\:w$    &     ~~~~$0.05$ ~~~~    &     ~~~~$0.1$~~~~       &     ~~~~$0.5$~~~~        &     ~~~~$0.9$~~~~   &        $0.99$           \\   \midrule[0.5pt] \vspace*{-8pt}  \\
                   0.01         &          8.89          &          8.76           &          5.36            &          0.94       &         0.09            \\
                    0.1         &          8.88          &          8.75           &          5.35            &          0.94       &         0.09            \\
                    0.5         &          8.64          &          8.51           &          5.21            &          0.91       &         0.09            \\
                      1         &          7.90          &          7.78           &          4.76            &          0.83       &         0.08            \\
                    1.9         &          5.32          &          5.24           &          3.21            &          0.56       &         0.05            \\
                  1.999         &          4.94          &          4.87           &          2.98            &          0.52       &         0.05            \\   \bottomrule[1.2px]
\end{tabular} \par  \vspace*{3pt}
     \centerline{(c) $\delta(\vartheta_1^{+}+\vartheta_1^{-})\,\varepsilon$ } \vspace*{10pt}
\end{minipage} \\

\begin{minipage}[t]{0.49\textwidth}
  \centering
\begin{tabular}{cccccccc} \toprule[1.2px]
      $\beta\:\backslash\:w$    &    ~~~~$ 0.05 $ ~~~~   &     ~~~~$0.1$~~~~       &     ~~~~$0.5$~~~~        &     ~~~~$0.9$~~~~   &        $0.99$                  \\   \midrule[0.5pt] \vspace*{-8pt}  \\
                   0.01         &          0.08          &          0.08           &     $\vartriangle$       &    $\vartriangle$   &    $\vartriangle$              \\
                    0.1         &          0.84          &          0.76           &          0.24            &    $\vartriangle$   &    $\vartriangle$              \\
                    0.5         &          4.06          &          3.65           &          1.09            &          0.10       &    $\vartriangle$              \\
                      1         &          7.42          &          6.60           &          1.72            &          0.12       &    $\vartriangle$              \\
                      5         &         13.01          &          9.28           &       $-$1.17            &       $-$0.38       &    $\vartriangle$              \\
                     10         &         10.19          &          4.47           &       $-$2.56            &       $-$0.51       &       $-$0.05                  \\   \bottomrule[1.2px]
\end{tabular} \par  \vspace*{3pt}
     \centerline{(d$_1$) $\delta(\vartheta_1^{+}-\vartheta_1^{-})\,\varepsilon$ $\left(\hat{\eta}=1.9\right)$ }
\end{minipage}
\begin{minipage}[t]{0.49\textwidth}
  \centering
\begin{tabular}{cccccccc} \toprule[1.2px]
 $\hat{\eta}\:\backslash\:w$    &    ~~~~$ 0.05 $ ~~~~   &     ~~~~$0.1$~~~~       &     ~~~~$0.5$~~~~        &     ~~~~$0.9$~~~~   &            $0.99$              \\   \midrule[0.5pt] \vspace*{-8pt}  \\
                   0.01         &          5.82          &          5.00           &          0.55            &        $-$0.11      &       $\vartriangle$           \\
                    0.1         &          5.82          &          5.00           &          0.56            &        $-$0.11      &       $\vartriangle$           \\
                    0.5         &          5.93          &          5.11           &          0.63            &        $-$0.09      &       $\vartriangle$           \\
                      1         &          6.26          &          5.44           &          0.88            &    $\vartriangle$   &       $\vartriangle$           \\
                    1.9         &          7.42          &          6.60           &          1.72            &           0.12      &       $\vartriangle$           \\
                  1.999         &          7.59          &          6.77           &          1.85            &           0.15      &       $\vartriangle$           \\   \bottomrule[1.2px]
\end{tabular} \par  \vspace*{3pt}
     \centerline{(d$_2$) $\delta(\vartheta_1^{+}-\vartheta_1^{-})\,\varepsilon$ $\left(\beta=1\right)$ } \vspace*{10pt}
\end{minipage}

\begin{minipage}[t]{0.49\textwidth}
  \centering
\begin{tabular}{cccccccc} \toprule[1.2px]
      $\beta\:\backslash\:w$    &    ~~~~$ 0.05 $ ~~~~   &     ~~~~$0.1$~~~~       &         ~~~~$0.5$~~~~           &        ~~~~$0.9$~~~~       &           $0.99$               \\   \midrule[0.5pt] \vspace*{-8pt}  \\
                   0.01         &          4.13          &          0.54           &        $\vartriangle$           &       $\vartriangle$       &       $\vartriangle$           \\
                    0.1         &          4.13          &          0.54           &        $\vartriangle$           &       $\vartriangle$       &       $\vartriangle$           \\
                    0.5         &          4.13          &          0.54           &        $\vartriangle$           &       $\vartriangle$       &       $\vartriangle$           \\
                      1         &          4.13          &          0.55           &        $\vartriangle$           &       $\vartriangle$       &       $\vartriangle$           \\
                      5         &          4.31          &          0.63           &        $\vartriangle$           &       $\vartriangle$       &       $\vartriangle$           \\
                     10         &          4.85          &          0.87           &        $\vartriangle$           &       $\vartriangle$       &       $\vartriangle$           \\   \bottomrule[1.2px]
\end{tabular} \par  \vspace*{3pt}
     \centerline{(e$_1$) $\delta(\vartheta_2^{+}+\vartheta_2^{-})\,\varepsilon^2$ $\left(\hat{\eta}=1.9\right)$ }
\end{minipage}
\begin{minipage}[t]{0.49\textwidth}
  \centering
\begin{tabular}{cccccccc} \toprule[1.2px]
 $\hat{\eta}\:\backslash\:w$    &    ~~~~$ 0.05 $ ~~~~   &     ~~~~$0.1$~~~~       &         ~~~~$0.5$~~~~           &        ~~~~$0.9$~~~~       &           $0.99$               \\   \midrule[0.5pt] \vspace*{-8pt}  \\
                   0.01         &          4.13          &          0.54           &        $\vartriangle$           &       $\vartriangle$       &       $\vartriangle$           \\
                    0.1         &          4.13          &          0.54           &        $\vartriangle$           &       $\vartriangle$       &       $\vartriangle$           \\
                    0.5         &          4.13          &          0.54           &        $\vartriangle$           &       $\vartriangle$       &       $\vartriangle$           \\
                      1         &          4.13          &          0.54           &        $\vartriangle$           &       $\vartriangle$       &       $\vartriangle$           \\
                    1.9         &          4.13          &          0.55           &        $\vartriangle$           &       $\vartriangle$       &       $\vartriangle$           \\
                  1.999         &          4.13          &          0.55           &        $\vartriangle$           &       $\vartriangle$       &       $\vartriangle$           \\   \bottomrule[1.2px]
\end{tabular} \par  \vspace*{3pt}
     \centerline{(e$_2$) $\delta(\vartheta_2^{+}+\vartheta_2^{-})\,\varepsilon^2$ $\left(\beta=1\right)$ } \vspace*{10pt}
\end{minipage}

\begin{minipage}[t]{0.49\textwidth}
  \centering
\begin{tabular}{cccccccc} \toprule[1.2px]
      $\beta\:\backslash\:w$    &    ~~~~$ 0.05 $ ~~~~   &     ~~~~$0.1$~~~~       &         ~~~~$0.5$~~~~           &        ~~~~$0.9$~~~~       &           $0.99$               \\   \midrule[0.5pt] \vspace*{-8pt}  \\
                   0.01         &     $\vartriangle$     &    $\vartriangle$       &        $\vartriangle$           &       $\vartriangle$       &       $\vartriangle$           \\
                    0.1         &     $\vartriangle$     &    $\vartriangle$       &        $\vartriangle$           &       $\vartriangle$       &       $\vartriangle$           \\
                    0.5         &        $-$0.15         &    $\vartriangle$       &        $\vartriangle$           &       $\vartriangle$       &       $\vartriangle$           \\
                      1         &        $-$0.29         &        $-$0.08          &        $\vartriangle$           &       $\vartriangle$       &       $\vartriangle$           \\
                      5         &        $-$1.46         &        $-$0.38          &        $\vartriangle$           &       $\vartriangle$       &       $\vartriangle$           \\
                     10         &        $-$2.91         &        $-$0.75          &        $\vartriangle$           &       $\vartriangle$       &       $\vartriangle$           \\   \bottomrule[1.2px]
\end{tabular} \par  \vspace*{3pt}
     \centerline{(f$_1$) $\delta(\vartheta_2^{+}-\vartheta_2^{-})\,\varepsilon^2$ $\left(\hat{\eta}=1.9\right)$ }
\end{minipage}
\begin{minipage}[t]{0.49\textwidth}
  \centering
\begin{tabular}{cccccccc} \toprule[1.2px]
 $\hat{\eta}\:\backslash\:w$    &    ~~~~$ 0.05 $ ~~~~   &     ~~~~$0.1$~~~~       &     ~~~~$0.5$~~~~           &        ~~~~$0.9$~~~~       &           $0.99$               \\   \midrule[0.5pt] \vspace*{-8pt}  \\
                   0.01         &        $-$0.29         &        $-$0.07          &    $\vartriangle$           &       $\vartriangle$       &       $\vartriangle$           \\
                    0.1         &        $-$0.29         &        $-$0.07          &    $\vartriangle$           &       $\vartriangle$       &       $\vartriangle$           \\
                    0.5         &        $-$0.29         &        $-$0.07          &    $\vartriangle$           &       $\vartriangle$       &       $\vartriangle$           \\
                      1         &        $-$0.29         &        $-$0.07          &    $\vartriangle$           &       $\vartriangle$       &       $\vartriangle$           \\
                    1.9         &        $-$0.29         &        $-$0.08          &    $\vartriangle$           &       $\vartriangle$       &       $\vartriangle$           \\
                  1.999         &        $-$0.29         &        $-$0.08          &    $\vartriangle$           &       $\vartriangle$       &       $\vartriangle$           \\   \bottomrule[1.2px]
\end{tabular} \par  \vspace*{3pt}
     \centerline{(f$_2$) $\delta(\vartheta_2^{+}-\vartheta_2^{-})\,\varepsilon^2$ $\left(\beta=1\right)$ }
\end{minipage}

\caption{The values (in units of $\mu$as) of $\delta\vartheta_1^{+}\varepsilon$, $\delta\vartheta_2^{+}\varepsilon^2$, $\delta(\vartheta_1^{+}\pm\vartheta_1^{-})\,\varepsilon$, and $\delta(\vartheta_2^{+}\pm\vartheta_2^{-})\,\varepsilon^2$ for various $\beta$ (or $\hat{\eta}$) and $w$. Here and thereafter, we use a triangle ``$\vartriangle$" to denote an absolute value which is smaller than $0.05\mu$as (the angular resolution of the NEAT mission), and fix our attention on the absolute values of the velocity effects when discussing their detectability. Additionally, we just discuss the cases of relativistic massive (and massless) particles, along with a mentioned rough lower limit $0.05$ for $w$ in geometrized units. }   \label{Table1}
\end{table}
\end{widetext}

\begin{figure*}
\centering
\begin{minipage}[b]{13cm}
\includegraphics[width=13cm]{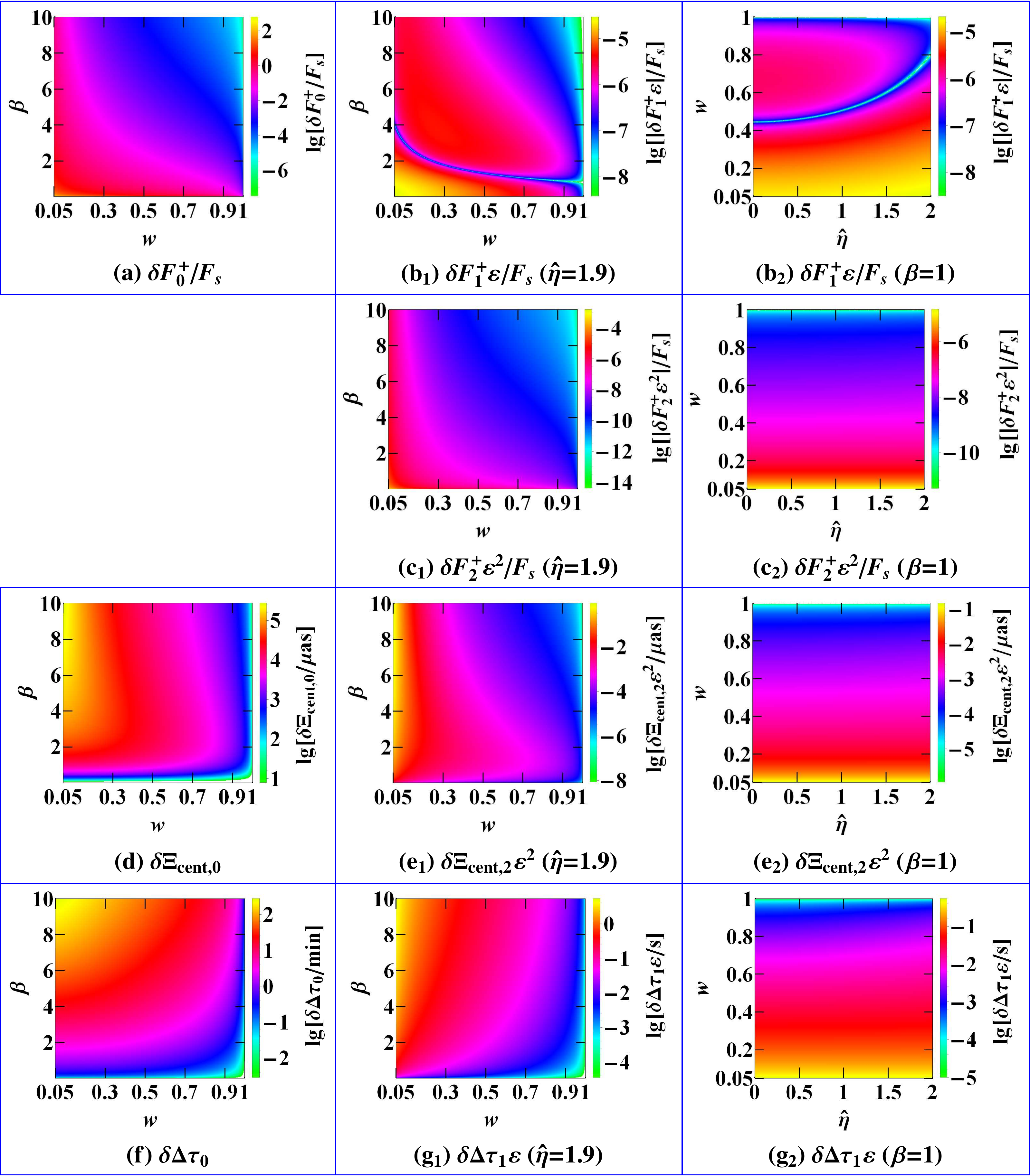}
\end{minipage}
\caption{The velocity effects on the zeroth-, first-, and second-order contributions to the normalized image flux, on the zeroth- and second-order contributions to the unscaled centroid position,
and on the zeroth- and first-order contributions to the unscaled (or practical) differential time delay of the light-source images, shown as the functions of $w$ and $\beta$ or $\hat{\eta}$ and $w$ (on the domain $\mathcal{D}_{\hat{\eta}1}$ or $\mathcal{D}_{\beta}$) in color-indexed form. The figures of $\delta F_0^{+}/F_{\text{s}},~\delta\Xi_{\text{cent,0}}$, and $\delta\Delta\tau_0$ are taken from~\cite{HL2022} for comparison.  } \label{Figure5}
\end{figure*}

\begin{figure*}
\centering
\begin{minipage}[b]{14cm}
\includegraphics[width=14cm]{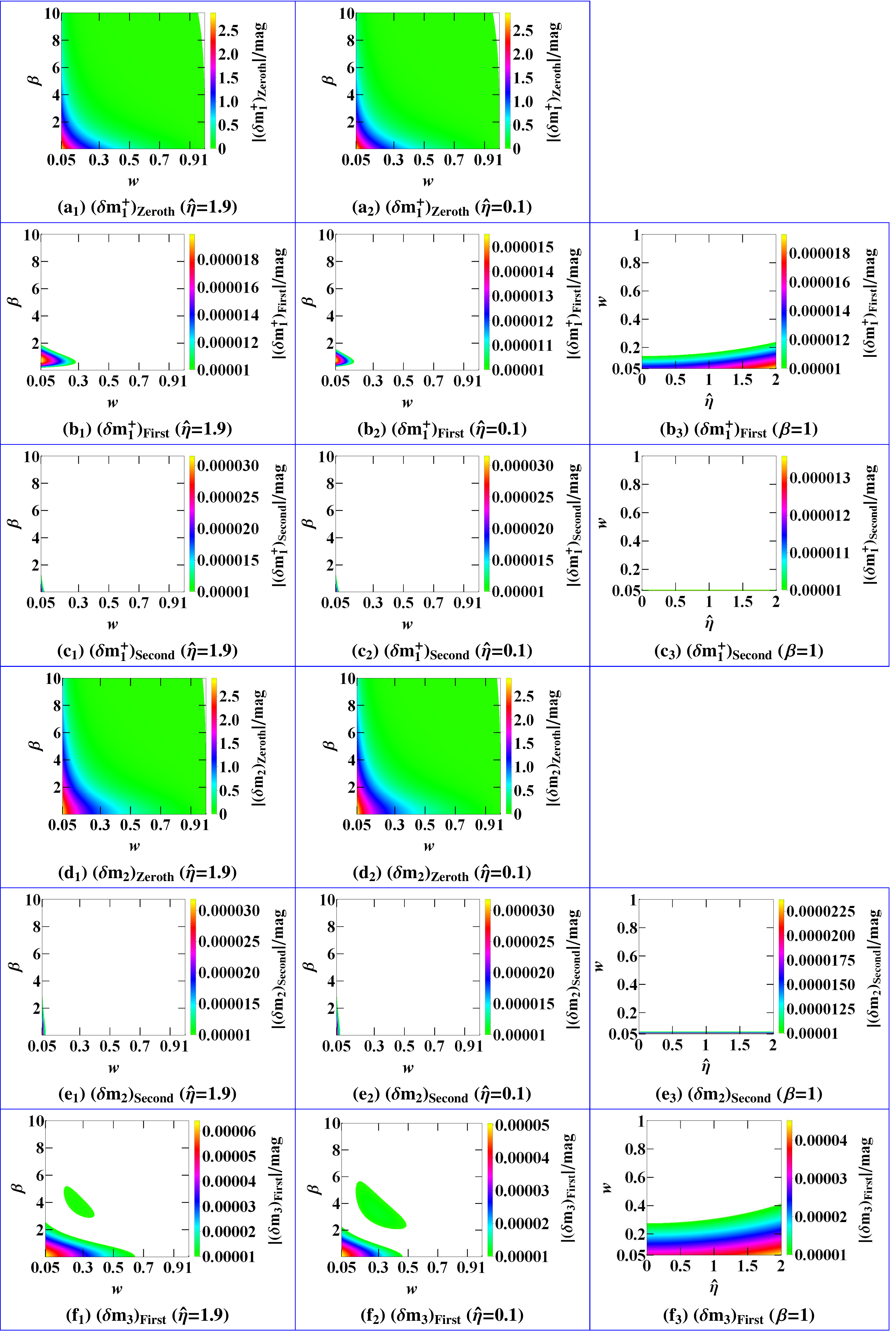}
\end{minipage}

\caption{The color-indexed magnitude-like differences caused by the velocity effects on the zeroth-, first-, and second-order contributions to $F^{+}$, on the zeroth- and second-order contributions to $F^{+}+F^{-}$,
and on the first-order contribution to $F^{+}-F^{-}$, plotted as the functions of $w$ and $\beta$ or $\hat{\eta}$ and $w$. In each of the figures here, we only show the part where the absolute value of a magnitude-like difference is not smaller than $10\mu$mag. As examples, we consider the cases of $\hat{\eta}=1.9$ and $\hat{\eta}=0.1$, respectively, when these magnitude-like differences act as the functions of $w$ and $\beta$. The case with $\beta=1$ is also given when $(\delta m_1^{+})_{\text{First}}$, $(\delta m_3)_{\text{First}}$, $(\delta m_1^{+})_{\text{Second}}$, and $(\delta m_2)_{\text{Second}}$ are the functions of $\hat{\eta}$ and $w$. } \label{Figure6}
\end{figure*}

We now turn our attention to the velocity effects on the individual fluxes and on the flux sum and difference relations of the images of the light source. The color-indexed velocity effects on the zeroth-, first-, and second-order contributions to the normalized flux of the light-source primary image are shown in Fig.~\ref{Figure5}, respectively. Notice that it is not necessary to plot the velocity effects on the sum and difference relations of the normalized fluxes of the light-source images because of Eqs. \eqref{SR2} - \eqref{SR5}. The values of $\delta F_1^{+}\varepsilon/F_{\text{s}}$ and $\delta F_2^{+}\varepsilon^2/F_{\text{s}}$ are presented in Tab.~\ref{Table2}, while the ones of $\delta F_0^{+}/F_{\text{s}}$ have been given in~\cite{HL2022}. In a similar way, we begin with the discussion of the changing trends of these velocity effects. The results given in Fig.~\ref{Figure5} and Tab.~\ref{Table2} suggest the velocity effects on the first-order normalized flux of the primary image and on the first-order normalized-flux difference of the images of the light source decrease monotonically on the domain $\left\{(w,~\beta,~\hat{\eta})\,|\,0.05\lesssim w<1,~0.01\leq\beta\lesssim0.8,~\hat{\eta}=1.9\right\}$ when increasing $w$. However, they first decrease to a minimum value and then increase to zero with the increase of $w$ from 0.05 to 1 on the domain $\left\{(w,~\beta,~\hat{\eta})\,|\,0.05\lesssim w<1,~0.8\lesssim\beta\leq10,~\hat{\eta}=1.9\right\}$ as well as on $\mathcal{D}_{\beta}$. Compared with them, the velocity effects $\delta F_2^{+}\varepsilon^2/F_{\text{s}}$ and $\delta(F_2^{+}+F_2^{-})\,\varepsilon^2/F_{\text{s}}$ increase with the increase of $w$ on both $\mathcal{D}_{\hat{\eta}1}$ and $\mathcal{D}_{\beta}$, in contrast to the behaviors of $\delta F_0^{+}/F_{\text{s}}$ and $\delta(F_0^{+}+F_0^{-})/F_{\text{s}}$~\cite{HL2022}. Then we estimate the possibilities of detecting these velocity effects, according to Fig.~\ref{Figure5}, Tab.~\ref{Table2}, and Fig.~\ref{Figure6} in which the magnitude-like differences induced by the velocity effects on the $i\,$th-order $\left(i\in\{0,~1,~2\}\right)$ flux of the light-source primary image, on the $j\,$th-order $\left(j\in\{0,~2\}\right)$ flux sum, and on the first-order flux difference of the light-source images are plotted in color-indexed form. In comparison with the relatively large possibility of detecting $\delta F_0^{+}/F_{\text{s}}$ and $\delta(F_0^{+}+F_0^{-})/F_{\text{s}}$~\cite{HL2022}, which can be indicated by the color-indexed results of $(\delta m_1^{+})_{\text{Zeroth}}$ and
$(\delta m_2)_{\text{Zeroth}}$ in Fig.~\ref{Figure6}, it is also likely to measure the velocity effects on the first-order normalized flux of the primary image and on the first-order normalized flux difference of the images of the light source by future multi-messenger emission detectors (or telescopes) whose flux resolution is approximately equal to or better than that of the nominal Kepler Mission (with a photometric precision of a few $\mu$mag~\cite{Koch2010,BK2018}). On most of the domain $\left\{(w,~\beta,~\hat{\eta})\,|\,0.05\lesssim w\lesssim 0.27,~0.2\lesssim\beta\lesssim1.8,~\hat{\eta}=1.9\right\}$ with a large bounce parameter, we find that the absolute value of the magnitude-like difference $(\delta m_1^{+})_{\text{First}}$ caused by $\delta F_1^{+}\varepsilon$ can be larger than $10\mu$mag and thus within the capability of these Kepler-level multi-messenger emission detectors. This is the case for $(\delta m_1^{+})_{\text{First}}$ on the main part of the domain $\left\{(w,~\beta,~\hat{\eta})\,|\,0.05\lesssim w\lesssim 0.17,~0.27\lesssim\beta\lesssim1.5,~\hat{\eta}=0.1\right\}$ (with a small bounce parameter) or of the domain $\left\{(w,~\beta,~\hat{\eta})\,|\,0.05\lesssim w\lesssim 0.23,~\beta=1,~0<\hat{\eta}<2\right\}$. Compared with the case of $(\delta m_1^{+})_{\text{First}}$, the magnitude-like difference $(\delta m_3)_{\text{First}}$ acting as a bivariate function of $\beta$ and $w$ possesses two separated and more evident regions where its absolute value is not smaller than $10\mu$mag for a given bounce parameter $\hat{\eta}\in(0,~2)$ (see, e.g., Fig.\,\ref{Figure6}\,(f$_1$) - (f$_2$) for details). It indicates a significantly larger chance to detect the velocity effect on the first-order difference relation of the normalized fluxes of the light-source images via the mentioned Kepler-level multi-messenger detectors. This conclusion can also be recognized from another perspective for $(\delta m_3)_{\text{First}}$ on the domain $\mathcal{D}_{\beta}$ with a given $\beta=1$, as shown in Fig.\,\ref{Figure6}\,(f$_3$). For example, $|(\delta m_3)_{\text{First}}|$ is about $12.3\mu$mag when $w=0.35,~\beta=1$, and $\hat{\eta}=1.8$ are given. It reaches $37.9\mu$mag for the case of $w=0.1,~\beta=1$, and $\hat{\eta}=1.9$. With respect to the velocity effects on the second-order normalized flux of the primary image and on the second-order normalized-flux sum of the images of the light source, it is only when $w$ takes a very limited value and $\beta$ has a small range that they may be measured by the future Kepler-level multi-messenger detectors, as shown in Fig.\,\ref{Figure6}\,(c$_1$) - (c$_3$) and Fig.\,\ref{Figure6}\,(e$_1$) - (e$_3$). For instance, with the assumption of $w=0.07,~\beta=0.1$, and $\hat{\eta}=0.1$, $(\delta m_1^+)_{\text{Second}}$ resulted from $\delta F_2^{+}\varepsilon^2$ and $(\delta m_2)_{\text{Second}}$ caused by $\delta(F_2^{+}+F_2^{-})\,\varepsilon^2$ have values $10.7\mu$mag and $11.7\mu$mag, respectively,
both of which are within the resolution capability of these Kepler-level multi-messenger detectors mentioned above.

\begin{widetext}

\begin{table}
\begin{minipage}[t]{0.48\textwidth}
  \centering
\begin{tabular}{cccccccc} \toprule[1.2px]
      $\beta\:\backslash\:w$    &    ~~~~$ 0.05 $ ~~~~   &      ~~~~$0.1$~~~~      &      ~~~~$0.5$~~~~      &     ~~~~$0.9$~~~~    &        $0.99$               \\   \midrule[0.5pt] \vspace*{-8pt}  \\
                   0.01         &  2.96$\times10^{-5}$   &   2.66$\times10^{-5}$   &   8.38$\times10^{-6}$   &        $\star$       &       $\star$           \\
                    0.1         &  2.94$\times10^{-5}$   &   2.65$\times10^{-5}$   &   8.29$\times10^{-6}$   &        $\star$       &       $\star$           \\
                    0.5         &  2.67$\times10^{-5}$   &   2.38$\times10^{-5}$   &   6.43$\times10^{-6}$   &        $\star$       &       $\star$           \\
                      1         &  2.03$\times10^{-5}$   &   1.74$\times10^{-5}$   &   2.34$\times10^{-6}$   &        $\star$       &       $\star$           \\
                      5         & $-$1.16$\times10^{-6}$ & $-$3.27$\times10^{-6}$  &  $-$2.02$\times10^{-6}$ &        $\star$       &       $\star$           \\
                     10         & $-$2.24$\times10^{-6}$ & $-$2.97$\times10^{-6}$  &         $\star$         &        $\star$       &       $\star$           \\   \bottomrule[1.2px]
\end{tabular} \par  \vspace*{3pt}
     \centerline{(a$_1$) $\delta F_1^{+}\varepsilon/F_{\text{s}}$ $\left(\hat{\eta}=1.9\right)$} \vspace*{10pt}
\end{minipage}  \hspace*{10pt}
\begin{minipage}[t]{0.48\textwidth}
  \centering
\begin{tabular}{cccccccc} \toprule[1.2px]
 $\hat{\eta}\:\backslash\:w$    &    ~~~~$ 0.05 $ ~~~~   &     ~~~~$0.1$~~~~       &     ~~~~$0.5$~~~~       &    ~~~~$0.9$~~~~    &        $0.99$              \\   \midrule[0.5pt] \vspace*{-8pt}  \\
                   0.01         &   1.58$\times10^{-5}$  &   1.29$\times10^{-5}$   &        $\star$          &       $\star$       &       $\star$           \\
                    0.1         &   1.58$\times10^{-5}$  &   1.30$\times10^{-5}$   &        $\star$          &       $\star$       &       $\star$           \\
                    0.5         &   1.61$\times10^{-5}$  &   1.33$\times10^{-5}$   &        $\star$          &       $\star$       &       $\star$           \\
                      1         &   1.71$\times10^{-5}$  &   1.42$\times10^{-5}$   &        $\star$          &       $\star$       &       $\star$           \\
                    1.9         &   2.03$\times10^{-5}$  &   1.74$\times10^{-5}$   &   2.34$\times10^{-6}$   &       $\star$       &       $\star$           \\
                  1.999         &   2.08$\times10^{-5}$  &   1.79$\times10^{-5}$   &   2.68$\times10^{-6}$   &       $\star$       &       $\star$           \\   \bottomrule[1.2px]
\end{tabular} \par  \vspace*{3pt}
     \centerline{(a$_2$) $\delta F_1^{+}\varepsilon/F_{\text{s}}$ $\left(\beta=1\right)$ }  \vspace*{10pt}
\end{minipage}
\begin{minipage}[t]{0.48\textwidth}
  \centering
\begin{tabular}{cccccccc} \toprule[1.2px]
      $\beta\:\backslash\:w$    &    ~~~~$ 0.05 $ ~~~~   &     ~~~~$0.1$~~~~      &     ~~~~$0.5$~~~~        &     ~~~~$0.9$~~~~    &        $0.99$           \\   \midrule[0.5pt] \vspace*{-8pt}  \\
                   0.01         & $-$1.47$\times10^{-3}$ & $-$1.92$\times10^{-4}$ &  $-$2.69$\times10^{-6}$  &        $\star$       &       $\star$           \\
                    0.1         & $-$1.47$\times10^{-4}$ & $-$1.92$\times10^{-5}$ &        $\star$           &        $\star$       &       $\star$           \\
                    0.5         & $-$2.93$\times10^{-5}$ & $-$3.84$\times10^{-6}$ &        $\star$           &        $\star$       &       $\star$           \\
                      1         & $-$1.46$\times10^{-5}$ & $-$1.91$\times10^{-6}$ &        $\star$           &        $\star$       &       $\star$           \\
                      5         & $-$2.80$\times10^{-6}$ &        $\star$         &        $\star$           &        $\star$       &       $\star$           \\
                     10         & $-$1.23$\times10^{-6}$ &        $\star$         &        $\star$           &        $\star$       &       $\star$           \\   \bottomrule[1.2px]
\end{tabular} \par  \vspace*{3pt}
     \centerline{(b$_1$) $\delta F_2^{+}\varepsilon^2/F_{\text{s}}$ $\left(\hat{\eta}=1.9\right)$} \vspace*{10pt}
\end{minipage} \hspace*{10pt}
\begin{minipage}[t]{0.48\textwidth}
  \centering
\begin{tabular}{cccccccc} \toprule[1.2px]
 $\hat{\eta}\:\backslash\:w$    &    ~~~~$ 0.05 $ ~~~~   &     ~~~~$0.1$~~~~      &     ~~~~$0.5$~~~~        &    ~~~~$0.9$~~~~    &        $0.99$              \\   \midrule[0.5pt] \vspace*{-8pt}  \\
                   0.01         & $-$1.46$\times10^{-5}$ & $-$1.91$\times10^{-6}$ &        $\star$           &       $\star$       &       $\star$           \\
                    0.1         & $-$1.46$\times10^{-5}$ & $-$1.91$\times10^{-6}$ &        $\star$           &       $\star$       &       $\star$           \\
                    0.5         & $-$1.46$\times10^{-5}$ & $-$1.91$\times10^{-6}$ &        $\star$           &       $\star$       &       $\star$           \\
                      1         & $-$1.46$\times10^{-5}$ & $-$1.91$\times10^{-6}$ &        $\star$           &       $\star$       &       $\star$           \\
                    1.9         & $-$1.46$\times10^{-5}$ & $-$1.91$\times10^{-6}$ &        $\star$           &       $\star$       &       $\star$           \\
                  1.999         & $-$1.46$\times10^{-5}$ & $-$1.91$\times10^{-6}$ &        $\star$           &       $\star$       &       $\star$           \\   \bottomrule[1.2px]
\end{tabular} \par  \vspace*{3pt}
     \centerline{(b$_2$) $\delta F_2^{+}\varepsilon^2/F_{\text{s}}$ $\left(\beta=1\right)$ }  \vspace*{10pt}
\end{minipage}
\begin{minipage}[t]{0.48\textwidth}
  \centering
\begin{tabular}{cccccccc} \toprule[1.2px]
      $\beta\:\backslash\:w$    &    ~~~~$ 0.05 $ ~~~~   &     ~~~~$0.1$~~~~       &     ~~~~$0.5$~~~~        &     ~~~~$0.9$~~~~    &           $0.99$          \\   \midrule[0.5pt] \vspace*{-8pt}  \\
                   0.01         &     $\vartriangle$     &     $\vartriangle$      &     $\vartriangle$       &    $\vartriangle$    &       $\vartriangle$      \\
                    0.1         &     $\vartriangle$     &     $\vartriangle$      &     $\vartriangle$       &    $\vartriangle$    &       $\vartriangle$      \\
                    0.5         &          0.07          &     $\vartriangle$      &     $\vartriangle$       &    $\vartriangle$    &       $\vartriangle$      \\
                      1         &          0.15          &     $\vartriangle$      &     $\vartriangle$       &    $\vartriangle$    &       $\vartriangle$      \\
                      5         &          0.65          &          0.12           &     $\vartriangle$       &    $\vartriangle$    &       $\vartriangle$      \\
                     10         &          0.94          &          0.10           &     $\vartriangle$       &    $\vartriangle$    &       $\vartriangle$      \\   \bottomrule[1.2px]
\end{tabular} \par  \vspace*{3pt}
     \centerline{(c$_1$) $\delta\Xi_{\text{cent,2}}\,\varepsilon^2$ $\left(\hat{\eta}=1.9\right)$} \vspace*{10pt}
\end{minipage} \hspace*{10pt}
\begin{minipage}[t]{0.48\textwidth}
  \centering
\begin{tabular}{cccccccc} \toprule[1.2px]
 $\hat{\eta}\:\backslash\:w$    &    ~~~~$ 0.05 $ ~~~~   &     ~~~~$0.1$~~~~       &     ~~~~$0.5$~~~~        &    ~~~~$0.9$~~~~    &            $0.99$           \\   \midrule[0.5pt] \vspace*{-8pt}  \\
                   0.01         &          0.14          &     $\vartriangle$      &     $\vartriangle$       &    $\vartriangle$   &        $\vartriangle$       \\
                    0.1         &          0.14          &     $\vartriangle$      &     $\vartriangle$       &    $\vartriangle$   &        $\vartriangle$       \\
                    0.5         &          0.14          &     $\vartriangle$      &     $\vartriangle$       &    $\vartriangle$   &        $\vartriangle$       \\
                      1         &          0.14          &     $\vartriangle$      &     $\vartriangle$       &    $\vartriangle$   &        $\vartriangle$       \\
                    1.9         &          0.14          &     $\vartriangle$      &     $\vartriangle$       &    $\vartriangle$   &        $\vartriangle$       \\
                  1.999         &          0.14          &     $\vartriangle$      &     $\vartriangle$       &    $\vartriangle$   &        $\vartriangle$       \\   \bottomrule[1.2px]
\end{tabular} \par  \vspace*{3pt}
     \centerline{(c$_2$) $\delta\Xi_{\text{cent,2}}\,\varepsilon^2$ $\left(\beta=1\right)$ }  \vspace*{10pt}
\end{minipage}

\begin{minipage}[t]{0.48\textwidth}
  \centering
\begin{tabular}{cccccccc} \toprule[1.2px]
      $\beta\:\backslash\:w$    &    ~~~~$ 0.05 $ ~~~~   &     ~~~~$0.1$~~~~       &     ~~~~$0.5$~~~~        &     ~~~~$0.9$~~~~    &           $0.99$               \\   \midrule[0.5pt] \vspace*{-8pt}  \\
                   0.01         &  5.31$\times10^{-3}$   &   2.52$\times10^{-3}$   &   2.30$\times10^{-4}$    & 1.81$\times10^{-5}$  &     1.55$\times10^{-6}$        \\
                    0.1         &          0.05          &          0.03           &   2.30$\times10^{-3}$    & 1.81$\times10^{-4}$  &     1.55$\times10^{-5}$        \\
                    0.5         &          0.27          &          0.13           &          0.01            & 9.07$\times10^{-4}$  &     7.76$\times10^{-5}$        \\
                      1         &          0.53          &          0.25           &          0.02            & 1.81$\times10^{-3}$  &     1.55$\times10^{-4}$        \\
                      5         &          2.65          &          1.26           &          0.12            & 9.07$\times10^{-3}$  &     7.76$\times10^{-4}$        \\
                     10         &          5.31          &          2.52           &          0.23            &         0.02         &     1.55$\times10^{-3}$        \\   \bottomrule[1.2px]
\end{tabular} \par  \vspace*{3pt}
     \centerline{(d$_1$) $\delta\Delta\tau_1\,\varepsilon$ $\left(\hat{\eta}=1.9\right)$} \vspace*{10pt}
\end{minipage} \hspace*{13pt}
\begin{minipage}[t]{0.48\textwidth}
  \centering
\begin{tabular}{cccccccc} \toprule[1.2px]
 $\hat{\eta}\:\backslash\:w$    &    ~~~~$ 0.05 $ ~~~~   &     ~~~~$0.1$~~~~       &     ~~~~$0.5$~~~~        &    ~~~~$0.9$~~~~    &            $0.99$              \\   \midrule[0.5pt] \vspace*{-8pt}  \\
                   0.01         &          0.53          &          0.25           &          0.02            & 9.67$\times10^{-4}$ &     7.16$\times10^{-5}$        \\
                    0.1         &          0.53          &          0.25           &          0.02            & 9.69$\times10^{-4}$ &     7.18$\times10^{-5}$        \\
                    0.5         &          0.53          &          0.25           &          0.02            & 1.03$\times10^{-3}$ &     7.74$\times10^{-5}$        \\
                      1         &          0.53          &          0.25           &          0.02            & 1.20$\times10^{-3}$ &     9.47$\times10^{-5}$        \\
                    1.9         &          0.53          &          0.25           &          0.02            & 1.81$\times10^{-3}$ &     1.55$\times10^{-4}$        \\
                  1.999         &          0.53          &          0.25           &          0.02            & 1.90$\times10^{-3}$ &     1.64$\times10^{-4}$        \\   \bottomrule[1.2px]
\end{tabular} \par  \vspace*{3pt}
     \centerline{(d$_2$) $\delta\Delta\tau_1\,\varepsilon$ $\left(\beta=1\right)$ }  \vspace*{10pt}
\end{minipage}

\caption{The values of $\delta F_1^{+}\varepsilon/F_{\text{s}}$ and $\delta F_2^{+}\varepsilon^2/F_{\text{s}}$, as well as those of $\delta\Xi_{\text{cent,2}}\,\varepsilon^2$ (in units of $\mu$as) and $\delta\Delta\tau_1\,\varepsilon$ (in units of seconds), for various $\beta$ (or $\hat{\eta}$) and $w$. A star ``$\star$" denotes the magnitude whose absolute value is less than $1.0\times10^{-6}$.}   \label{Table2}
\end{table}
\end{widetext}

We finally discuss the velocity effects on the magnification centroid and on the differential time delay of the images of the light source. On the one hand Figure~\ref{Figure5} shows the color-indexed velocity effects on the zeroth- and second-order contributions to the unscaled centroid position, as well as those on the zeroth- and first-order contributions to the unscaled differential time delay, as the bivariate functions of $\beta$ (or $\hat{\eta}$) and $w$. Additionally, the values of $\delta\Xi_{\text{cent,2}}\,\varepsilon^2$ and $\delta\Delta\tau_1\,\varepsilon$ are given in Tab.~\ref{Table2}. Similar to the behavior of $\delta\Xi_{\text{cent,}0}$ or $\delta\Delta\tau_0$, Figure~\ref{Figure5} implies that both $\delta\Xi_{\text{cent,}2}\varepsilon^2$ and $\delta\Delta\tau_1\varepsilon$ decrease monotonically when increasing $w$ on $\mathcal{D}_{\hat{\eta}1}$ (with a fixed $\beta$) or on $\mathcal{D}_{\beta}$ (with a fixed $\hat{\eta}$). On the other hand, we consider the possibilities of their astronomical detections. In comparison with the good chance to observe $\delta\Xi_{\text{cent,}0}$, it is not until $w$ takes a very limited value while $\beta$ takes a relatively large value that there is a chance to detect $\delta\Xi_{\text{cent,}2}\varepsilon^2$ via the NEAT-level multi-messenger detectors mentioned above, according to the results on the domains $\mathcal{D}_{\hat{\eta}1}$ and $\mathcal{D}_{\beta}$ shown in Fig.~\ref{Figure5} and Tab.~\ref{Table2}. For instance, if $w=0.07,~\beta=7$, and $\hat{\eta}=1.9$ are assumed, then the value of $\delta\Xi_{\text{cent,}2}\varepsilon^2$ is about 0.35$\mu$as, which is seven times the precision of those NEAT-level multi-messenger detectors. Unlike the case of $\delta\Xi_{\text{cent,}2}\varepsilon^2$, Figure~\ref{Figure5} and Table~\ref{Table2} suggest a large possibility to detect the velocity effect on the first-order unscaled differential time delay of light signals in current resolution, whether on $\mathcal{D}_{\hat{\eta}1}$ or on $\mathcal{D}_{\beta}$. For example, even for an ordinary massive neutrino with $w\approx0.999999$~\cite{Adam2012,Adamson2015} serving as the test particle, the value of $\delta\Delta\tau_1\varepsilon$ can reach about $9.2\times10^{-9}$s which is larger than four times the current time resolution ($\sim2$ns) of the IceCube neutrino detector~\cite{IceCube2006,AH2018} and is about nine times the time resolution ($\sim1$ns) of the ARGO-YBJ air shower detector~\cite{Aielli2006,Bolo2020}, if $\beta=1$ and $\hat{\eta}=1$ are assumed.

\begin{figure*}
\centering
\begin{minipage}[b]{13.cm}
\includegraphics[width=13.cm]{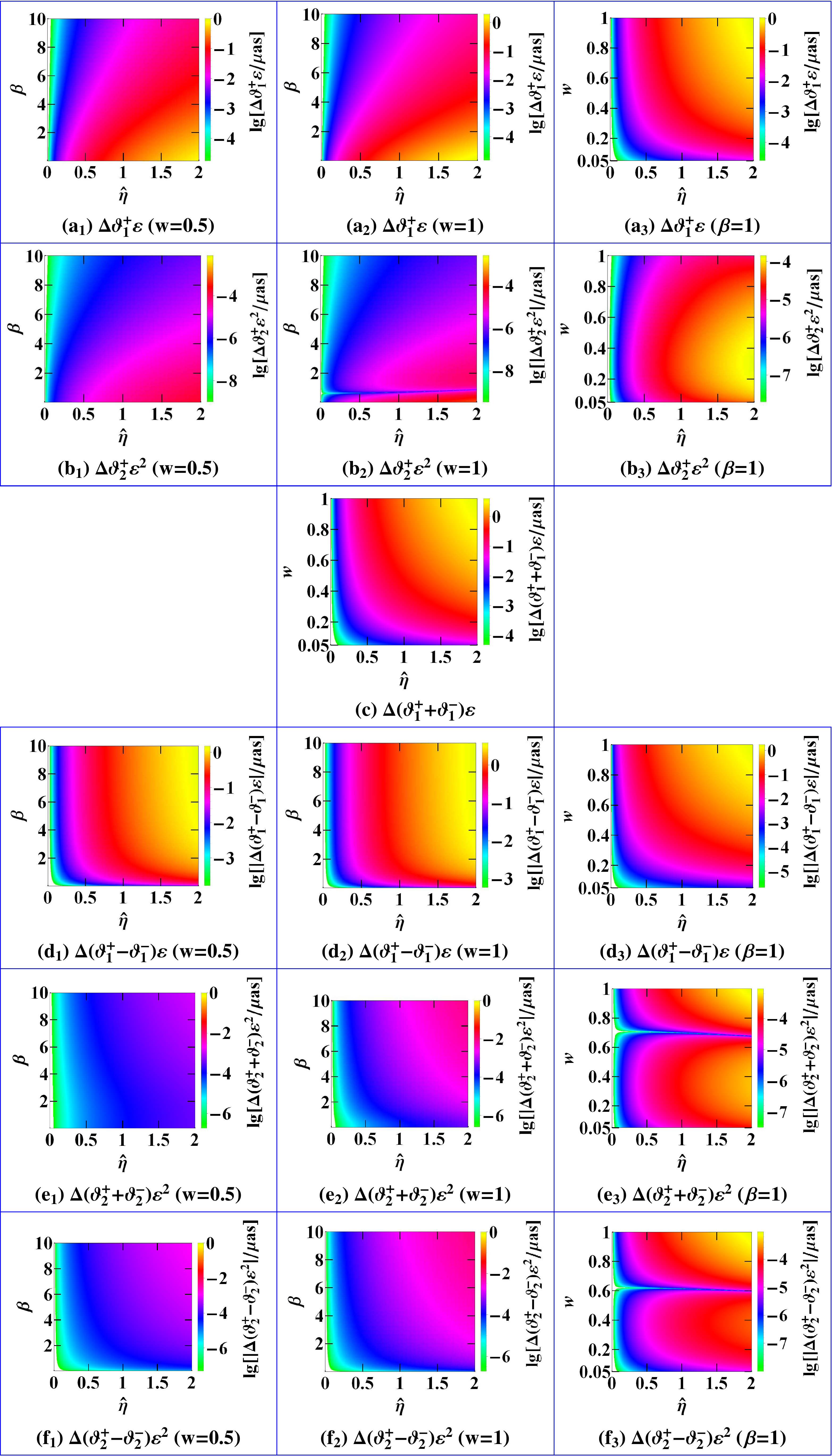}
\end{minipage}

\caption{The color-indexed bounce effects on the first- and second-order contributions to the unscaled primary-image position and to the unscaled positional sum and difference relations plotted as the functions of $\hat{\eta}$ and $\beta$ or $\hat{\eta}$ and $w$. } \label{Figure7}
\end{figure*}

\begin{figure*}
\centering

\begin{minipage}[b]{13cm}
\includegraphics[width=13cm]{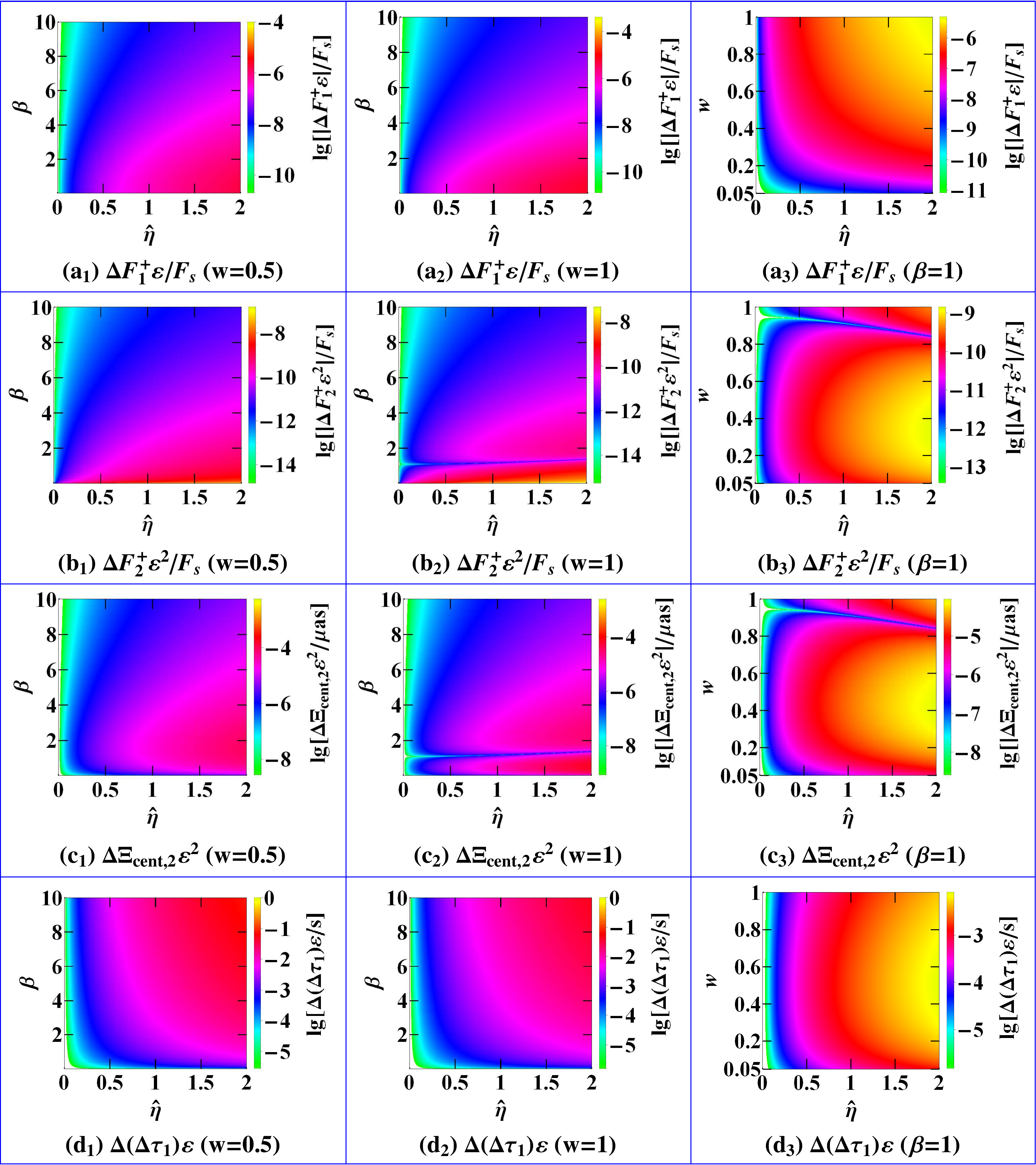}
\end{minipage}

\caption{The color-indexed bounce effects on the first- and second-order contributions to the normalized flux of the primary particle-source image, on the second-order unscaled centroid, and on the first-order unscaled differential time delay, shown as the functions of $\hat{\eta}$ and $w$ or $\hat{\eta}$ and $\beta$. }  \label{Figure8}
\end{figure*}

\begin{figure*}
\centering
\begin{minipage}[b]{14cm}
\includegraphics[width=14cm]{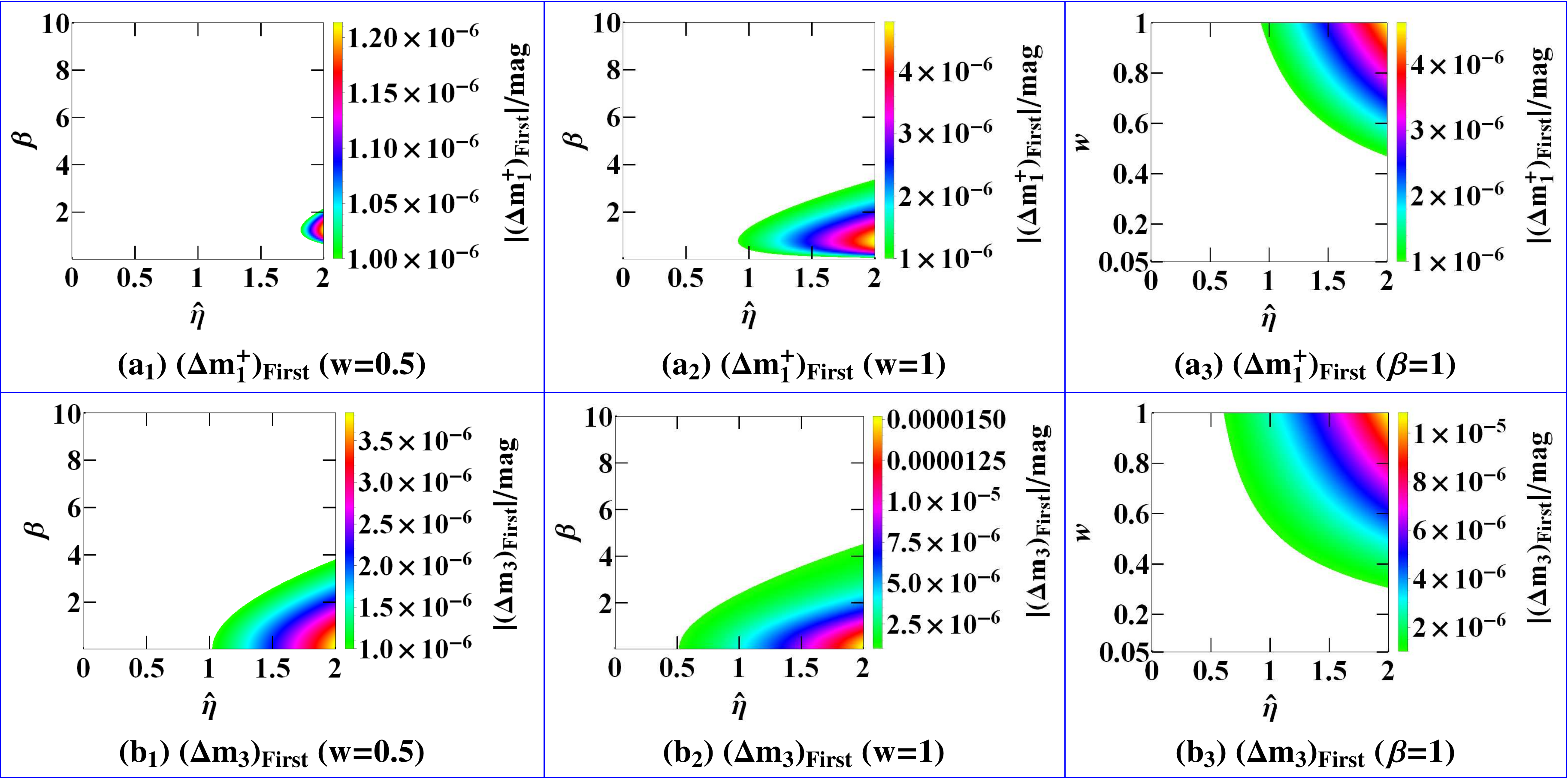}
\end{minipage}

\caption{The color-indexed magnitude-like differences caused by the bounce effects on the first-order flux of the primary image of the particle source and on the first-order flux difference of the particle-source images, plotted as the functions of $\beta$ (or $w$) and $\hat{\eta}$. Here, only the part where the absolute value of a magnitude-like difference is not smaller than $1\mu$mag is shown in each figure.  } \label{Figure9}
\end{figure*}

\subsection{Bounce effects on the practical lensing observables}
Next we focus on discussing the bounce-induced effects on the Schwarzschild lensing properties of the images of the massive-particle source. Firstly, Figure~\ref{Figure7} shows the bounce effects on the first- and second-order contributions to the unscaled position of the positive-parity image and to the unscaled positional relations of the images of the particle source in color-indexed form on the domains $\mathcal{D}_{w1},~\mathcal{D}_{w2}$, and $\mathcal{D}_{\beta}$, respectively. Compared with the behaviors of the velocity effects, these bounce effects exhibit a more complex dependence on the variables. It is found that the bounce effects on the first-order primary-image position and on the first-order positional sum relation increase monotonically with the increase of $\hat{\eta}$, for a given $\beta$ on any of the domains $\mathcal{D}_{w1}$ and $\mathcal{D}_{w2}$ or for a given $w$ on $\mathcal{D}_{\beta}$. This tendency applies to the bounce effect on the second-order image position on the domain $\left\{(w,~\beta,~\hat{\eta})\,|\,0.05\lesssim w\lesssim0.96,~\beta=1,~0<\hat{\eta}<2\right\}$ or $\left\{(w,~\beta,~\hat{\eta})\,|\,w=w_1,~0.01\leq\beta\leq10,~0<\hat{\eta}<2\right\}$, where $w_1$ is a given constant with a range $0.05\lesssim w_1\lesssim0.7$, as well as to that on the second-order positional sum on $\mathcal{D}_{w1}$ and to that on the second-order positional difference on $\mathcal{D}_{w2}$. However, for a given $\beta$ or $w$, the bounce effects on the first-order positional difference within $\mathcal{D}_{w1},~\mathcal{D}_{w2}$, or $\mathcal{D}_{\beta}$, on the second-order positional sum within $\mathcal{D}_{w2}$, and on the second-order positional difference within $\mathcal{D}_{w1}$, have a reverse tendency. Moreover, we find that the bounce effects on the second-order image position and on the second-order positional relations appear strange in some circumstances. Apart from experiencing a first increase and a later decrease with increasing $\hat{\eta}$ for a given $w$ on the domain $\left\{(w,~\beta,~\hat{\eta})\,|\,0.96\lesssim w\leq1,~\beta=1,~0<\hat{\eta}<2\right\}$, $\Delta\vartheta_2^{+}\varepsilon^2$ on $\left\{(w,~\beta,~\hat{\eta})\,|\,w=w_2,~0.01\leq\beta\leq10,~0<\hat{\eta}<2\right\}$ where the given constant $w_2$ takes a range $0.7\lesssim w_2\leq1$, has three different behaviors to which three different sub-ranges of the given parameter $\beta$ correspond. Similarly, if $\hat{\eta}$ increases, then $\Delta(\vartheta_2^{+}+\vartheta_2^{-})\varepsilon^2$ with a given $w$ will increase on the domain $\left\{(w,~\beta,~\hat{\eta})\,|\,0.05\lesssim w\lesssim0.66,~\beta=1,~0<\hat{\eta}<2\right\}$, decrease on the domain $\left\{(w,~\beta,~\hat{\eta})\,|\,0.7\lesssim w\leq1,~\beta=1,~0<\hat{\eta}<2\right\}$, and increase firstly and then decrease on the domain $\left\{(w,~\beta,~\hat{\eta})\,|\,0.66\lesssim w\lesssim0.7,~\beta=1,~0<\hat{\eta}<2\right\}$. In comparison with the bounce effect on the second-order positional sum, $\Delta(\vartheta_2^{+}-\vartheta_2^{-})\varepsilon^2$ experiences an inverse tendency of change on the domains $\left\{(w,~\beta,~\hat{\eta})\,|\,0.05\lesssim w\lesssim0.59,~\beta=1,~0<\hat{\eta}<2\right\}$, $\left\{(w,~\beta,~\hat{\eta})\,|\,0.62\lesssim w\leq1,~\beta=1,~0<\hat{\eta}<2\right\}$, and $\left\{(w,~\beta,~\hat{\eta})\,|\,0.59\lesssim w\lesssim0.62,~\beta=1,~0<\hat{\eta}<2\right\}$, respectively. Secondly, Figure~\ref{Figure7} indicates that there is a relatively large possibility to measure the bounce-induced effects on the first-order unscaled angular position of the primary image and on the first-order unscaled positional sum and difference relations of the images of the particle source by future high-accuracy particle detectors (or instruments) whose angular resolution is approximately equal to or better than that of the NEAT mission. Concretely, we find that the value of $\Delta\vartheta_1^+\varepsilon$ can be larger than $0.05\mu$as on about one out of two of the domain $\left\{(w,~\beta,~\hat{\eta})\,|\,w=0.5,~0.01\leq\beta\lesssim8.4,~0.5\lesssim\hat{\eta}<2\right\}$ $($or the domain $\left\{(w,~\beta,~\hat{\eta})\,|\,w=1,~0.01\leq\beta\lesssim8.7,~0.32\lesssim\hat{\eta}<2\right\})$ for a relativistic massive particle (or a photon) being the test particle. It is also true for $\Delta\vartheta_1^+\varepsilon$ on most of the domain $\left\{(w,~\beta,~\hat{\eta})\,|\,0.12\lesssim w\leq1,~\beta=1,~0.43\lesssim\hat{\eta}<2\right\}$. The maximum value of $\Delta\vartheta_1^+\varepsilon$ on those three domains reach about $0.79\mu$as, $1.97\mu$as, and $1.10\mu$as, respectively. Additionally, it is interesting to find that the value of $\Delta(\vartheta_1^{+}+\vartheta_1^{-})\varepsilon$ on most of the domain $\left\{(w,~\beta,~\hat{\eta})\,|\,0.38\lesssim w\leq1,~1\lesssim\hat{\eta}<2\right\}$, as well as the absolute value of $\Delta(\vartheta_1^{+}-\vartheta_1^{-})\varepsilon$ on almost all of a larger domain
$\left\{(w,~\beta,~\hat{\eta})\,|\,w=1,~0.53\lesssim\beta\leq10,~1.1\lesssim\hat{\eta}<2\right\}$, can exceed the angular resolution of these NEAT-level particle detectors apparently, and that their maximum absolute values are about $3.97\mu$as and $3.89\mu$as respectively. Furthermore, on almost all of the domain $\left\{(w,~\beta,~\hat{\eta})\,|\,w=0.5,~0.1\lesssim\beta\lesssim10,~0.37\lesssim\hat{\eta}<2\right\}$ and on most of $\left\{(w,~\beta,~\hat{\eta})\,|\,0.22\lesssim w\leq1,~\beta=1,~0.34\lesssim\hat{\eta}<2\right\}$, the absolute value of $\Delta(\vartheta_1^{+}-\vartheta_1^{-})\varepsilon$ is also larger than the astrometric precision of the mentioned NEAT-level particle detectors. As to the bounce-induced effects on the second-order unscaled primary-image position and on the second-order unscaled positional sum and difference, we conclude that there is not any possibility to detect them within the capability of these NEAT-level particle detectors. For instance, the largest absolute values of $\Delta\vartheta_2^{+}\varepsilon^2$ on $\mathcal{D}_{w1}$ and $\mathcal{D}_{w2}$ are about $1.7\times10^{-4}\mu$as and $2.2\times10^{-4}\mu$as respectively, two orders of magnitude less than the angular precision of these NEAT-level particle detectors. In the best case, $\Delta(\vartheta_2^{+}+\vartheta_2^{-})\varepsilon^2$ (or $|\Delta(\vartheta_2^{+}-\vartheta_2^{-})\varepsilon^2|$) on $\mathcal{D}_{w1}$ is about 40 times less than $0.05\mu$as.

We are now in a position to consider the bounce-induced effects on the fluxes and on the flux sum and difference relations of the images of the particle source. Figure~\ref{Figure8} presents the bounce effects on the first- and second-order contributions to the normalized flux of the primary image of the particle source in color-indexed form. And the corresponding color-indexed magnitude-like differences induced by the bounce effects on the first-order primary-image flux and on the first-order flux difference for the case of massive particles being the test particles are shown in Fig.~\ref{Figure9}. We notice that the bounce effects on the first-order primary-image flux and on the first-order flux difference decrease with increasing $\hat{\eta}$ monotonically, for a given $w$ on $\mathcal{D}_{\beta}$ or for a given $\beta$ on any of the domains $\mathcal{D}_{w1}$ and $\mathcal{D}_{w2}$. This tendency also applies to the bounce effect on the second-order primary-image flux on $\mathcal{D}_{w1}$ with a given $\beta$. However, the behavior of $\Delta F_2^+\varepsilon^2/F_s$ on $\mathcal{D}_{w2}$ appears more complex. Specifically, with a fixed source position and the increase of $\hat{\eta}$, it always increases on the domain $\left\{(w,~\beta,~\hat{\eta})\,|\,w=1,~0.01\leq\beta\lesssim1.1,~0\lesssim\hat{\eta}<2\right\}$, decreases on $\left\{(w,~\beta,~\hat{\eta})\,|\,w=1,~1.5\lesssim\beta\leq10,~0\lesssim\hat{\eta}<2\right\}$, and experiences a first decrease and a later increase on $\left\{(w,~\beta,~\hat{\eta})\,|\,w=1,~1.1\lesssim\beta\lesssim1.5,~0\lesssim\hat{\eta}<2\right\}$. Additionally, $\Delta F_2^+\varepsilon^2/F_s$ (or $\Delta(F_2^{+}+F_2^{-})\varepsilon^2$) on the corresponding sub-ranges of $\mathcal{D}_{\beta}$ has a reverse tendency of change partially. Moreover, Figures~\ref{Figure8} - \ref{Figure9} indicate that there is a small chance to detect the bounce effect on the first-order normalized flux difference of the particle-source images by future particle emission detectors whose flux resolution is approximately equal to (or better than) that of the Kepler Mission (with a photometric precision of a few $\mu$mag), since the value of the magnitude-like difference $(\Delta m_3)_{\text{First}}$ caused by it on most of the domain $\left\{(w,~\beta,~\hat{\eta})\,|\,w=1,~0.01\leq\beta\lesssim1.1,~1.63\lesssim\hat{\eta}<2\right\}$ or on half of $\left\{(w,~\beta,~\hat{\eta})\,|\,0.94\lesssim w<1,~\beta=1,~1.94\lesssim\hat{\eta}<2\right\}$ is larger than $10\mu$mag. We also find that the possibility to detect the bounce effect on the first-order normalized flux of the primary particle-source image by the capability of these Kepler-level particle detectors is very small or even non-existent, with the consideration of the maximum value $4.79\mu$mag of the magnitude-like difference $(\Delta m_1^{+})_{\text{First}}$ on the domains $\mathcal{D}_{w1},~\mathcal{D}_{w2}$, and $\mathcal{D}_{\beta}$. The results also show that the values of the bounce effects on the second-order normalized flux of the particle-source image and on the second-order normalized flux sum of the images of the particle source are so small that they are far beyond the capability of the mentioned Kepler-level particle detectors.

Eventually, we move on to the bounce effects on the centroid position and on the differential time delay in the Schwarzschild lensing phenomenon of massive particles. Figure~\ref{Figure8} exhibits the color-indexed bounce effects on the second-order unscaled centroid and on the first-order unscaled differential time delay. It indicates that the bounce effect on the second-order unscaled centroid on $\mathcal{D}_{w1}$ increases with increasing $\hat{\eta}$ for a given $\beta$, which also applies to the bounce effect on the first-order unscaled differential time delay for a given $w$ on $\mathcal{D}_{\beta}$ or a fixed $\beta$ on one of the domains $\mathcal{D}_{w1}$ and $\mathcal{D}_{w2}$. With different sub-ranges of $\beta$ (or $w$), the change trend of $\Delta\Xi_{\text{cent,2}}\varepsilon^2$ is contrary to that of $\Delta F_2^+\varepsilon^2/F_s$ when they are on the domain $\mathcal{D}_{w2}$ (or $\mathcal{D}_{\beta}$). As regards the possibility to detect them, we conclude that it is only by future high-accuracy particle detectors whose angular resolution is much higher than the one of the NEAT that it is possible to measure the bounce effect on the second-order unscaled centroid. However, a large possibility to detect the bounce effect on the first-order unscaled differential time delay in current resolution is shown by the results in Fig.~\ref{Figure8}, no matter which domain (among $\mathcal{D}_{w1},~\mathcal{D}_{w2}$, and $\mathcal{D}_{\beta}$) it lies on. Although the source position and the bounce parameter take very small values and the initial velocity of the massive particle takes a low relativistic value, the value of $\Delta(\Delta\tau_1)\varepsilon$ may still exceed the present precision of measuring the differential time delay remarkably. For example, if $w=0.08$, $\hat{\eta}=0.01$, and $\beta=0.1$ are fixed, then $\Delta(\Delta\tau_1)\varepsilon\approx5.46\,$ns, which is evidently larger than the time resolution of the ARGO-YBJ detector.

\section{Summary and discussion} \label{sect7}
In summary, the gravitational lensing of a massive neutral particle with a relativistic initial velocity in the black-bounce-Schwarzschild black hole spacetime have been studied beyond the weak-deflection limit. Based on the standard form of the black-bounce-Schwarzschild metric, we have adopted a classical approach to derive the gravitational deflection of the massive particle propagating in the equatorial plane of the lens in the 3PM approximation. The resulting deflection angle extends the results in the previous literatures, and has been utilized to solve the popular Virbhadra-Ellis lens equation for analyzing the observable properties of the primary and secondary images of a point source of the massive particle. We have obtained the weak-field expressions for the positions, magnifications, and gravitational time delays of the individual images, and those for the position and magnification relations (including the total magnification), the magnification centroid, and the differential time delay between the images. The velocity-induced effects, which is caused by the deviation of the initial velocity of the massive particle from the speed of light, on the measurable lensing properties of the images of a point-like light source in the black-bounce-Schwarzschild geometry have been probed. We have also discussed the effects induced by the bounce parameter of the spacetime on the Schwarzschild lensing observables of the images of the particle source. Serving as an application of our analytical results, the Galactic Center supermassive black hole, Sgr A$^{\ast}$, has been assumed as the black-bounce-Schwarzschild lens. Under this astrophysical scenario, we have assessed the possibilities to detect the new velocity-induced and bounce-induced effects on the practical lensing observables, along with the analysis of the dependence of these two effects on the source position, the initial velocity of the particle, and the bounce parameter.

Similar to the case of Kerr-Newman black hole lensing of massive particles~\cite{HL2022}, we find that the decrease of the particle's initial velocity leads to a monotonous increase of many components of the velocity effects on the black-bounce-Schwarzschild lensing observables of the images of the light source, for a given bounce parameter and a fixed scaled source position on proper domains respectively. These components include but are not limited to the velocity effects on the first- and second-order contributions to the primary-image position, on the first- and second-order sum relations of the unscaled positions, on the first-order normalized primary-image flux, on the first-order normalized-flux difference, on the second-order practical magnification centroid, and on the first-order unscaled differential time delay. It means that under the same conditions, the black-bounce-Schwarzschild lensing observables of the images of a massive-particle source become more apparent than those of the images of a light source for quite a few scenarios. It also implies that there may be some chances to detect the velocity effects on the observable features of the lensed images of the light source in the black-bounce-Schwarzschild spacetime, if the future (or even the current) techniques and avenues are enough for their astronomical measurements. Our results suggest a relatively large possibility of the detection of the velocity effects on the first-order unscaled differential time delay of the light-source images with current time resolution in substantial circumstances. With a tighter restriction on the domains of the parameters, it is also possible to measure the velocity effects on the first-order unscaled primary-image position, on the first-order normalized flux of the light-source primary image, and on the first-order practical image position sum, as well as those on the first-order difference relations of the unscaled positions and of the normalized fluxes of the light-source images, in the capability of future multi-messenger detectors whose angular resolution is similar to that of the NEAT Mission and whose flux resolution is similar to the one of the Kepler Mission. It is interesting to note that there is a rather small window of opportunity for detecting the velocity effects on the second-order unscaled image position, on the second-order normalized flux of the light-source primary image, on the second-order practical positional sum and difference relations, on the second-order normalized-flux sum, and on the second-order unscaled centroid position by these future multi-messenger detectors. Additionally, the measurements of the other velocity-effect components are beyond the resolution capability of these multi-messenger detectors.

We also find that some bounce effects on the measurable image properties in the Schwarzschild lensing of massive particles increase monotonically with the increase of the bounce parameter, when the initial velocity of the particle and the particle-source position are fixed respectively on proper ranges or sub-ranges. They contain not only the bounce effects on the first- and second-order contributions to the angular position of the primary image, but also the bounce effects on the first-order sum of the image positions and on the second-order positional sum and difference relations, in company with those on the second-order unscaled centroid and on the first-order unscaled differential time delay. A reverse dependence on the bounce parameter applies to the bounce effects on the first-order positional difference, the bounce effects on the first- and second-order contributions to the primary-image flux, and so on, when $w$ and $\beta$ are given on proper domains respectively. All of them indicate a potential chance to observe some of these bounce effects and thus to distinguish between the Schwarzschild and black-bounce-Schwarzschild black holes via the lensing properties related to timelike signals, once some critical value of the bounce parameter is exceeded. We find a relatively large possibility to measure the bounce effects on the first-order unscaled differential time delay of the particle-source images in current or near future astronomical accuracy. Moreover, it is possible to detect the bounce effects on the first-order unscaled primary-image position and on the first-order unscaled positional sum and difference of the images of the particle source by future particle detectors whose angular resolution is similar to that of the NEAT mission. And there is also a glimmer of hope to measure the bounce effect on the first-order normalized image-flux difference relation via future particle detectors whose flux resolution is similar to that of the Kepler Mission, which may even apply to the bounce effect on the first-order normalized primary-image flux. Additionally, our results indicate that it is not until via future high-accuracy particle detectors whose angular and flux resolution capabilities are better than NEAT's and Kepler's performances respectively that we may detect the bounce effects on the second-order unscaled position and the second-order normalized flux of the particle-source primary image, on the second-order unscaled positional sum and difference, on the second-order normalized-flux sum, and on the second-order unscaled magnification centroid of the images of the particle source.

As to the energy (or velocity) distribution of particle emission, we know gravitational time delay makes the massive particles emitted at the same time with different initial velocities arrive at the observer at different times, which leads to a result that the observed energy distribution is different from the emitted one. In this work, we only calculate the gravitational time delay for the massive particles with an arbitrary initial velocity, without considering a specific energy distribution. In our future work, we plan to study the relations between the observed energy distribution and the emitted one for various initial velocity distributions when the massive particles emitted by the source pass by a black-bounce-Schwarzschild black hole.

Finally, we must admit that some challenges and difficulties are to be overcome in current or future measurements of the gravitational lensing of massive particles and the related velocity-induced and bounce-induced effects. Nevertheless, it should be recognized that the desired information and features of the lenses and the particle sources (e.g., the sources of particle dark matter~\cite{Feng2010} or cosmic-ray particles~\cite{Anderson1932,Fermi1949}) may be revealed through the lensing effects of timelike signals. With the rapid progress made in the theoretical and observational investigations of the weak- and strong-field gravitational lensing phenomena (see, e.g.,~\cite{Perlick2004,VK2008,Virbha2009,LPH2012,CJ2015J,ZX2017E,ZJC2017,ZX2017P,CS2017,HL2017a,CX2018,CZJ2018,JO2018P,LX2019M,WSX2019,TBRE2020,FP2021,LX2021,
GX2021,LFZMH2021,Arakida2021,Tsupko2021,HLL2021a,HLL2021b,GX2022,JRPO2022,Virbhadra2022,Virb2022b,PPOD2023,WCJ2023,GLZ2023,SPVR2023,LZLCW2023,SVP2023,PO2023,MGKS2023}) and of the muiti-messenger synergic observations (see, for instance,~\cite{BT2017,MFHM2019,Keivani2018,IceCube2018,Huerta2019,QJFZZ2021}) over the last decades, it seems reasonable to expect a bright prospect for the gravitational lensing of massive particles.

\begin{acknowledgments}
We thank the anonymous reviewers very much for their useful and important comments on improving the quality of this work. G.H. is supported partially by the National Natural Science Foundation of China (Grant No. 12205139) and the Natural Science Foundation of Hunan Province (Grant No. 2022JJ40347). Y.X. is funded by the National Natural Science Foundation of China (Grant Nos. 12273116 and 62394351), the science research grants from the China Manned Space Project (Grant Nos. CMS-CSST-2021-A12 and CMS-CSST-2021-B10) and the Opening Project of National Key Laboratory of Aerospace Flight Dynamics of China (Grant No. KGJ6142210220201). W.L. was supported in part by the National Natural Science Foundation of China (Grant No. 11973025).
\end{acknowledgments}

\end{document}